\documentclass{pasj00}

\begin{document}
\SetRunningHead{Y. Takeda \& S. UeNo}{Radial--tangential macroturbulence model 
for the spectral line broadening of solar-type stars}
\Received{2016/12/26}
\Accepted{2017/03/07}

\title{Does the radial--tangential macroturbulence model adequately 
describe the spectral line broadening of solar-type stars? 
\thanks{Based on data collected by the Domeless Solar Telescope at 
Hida Observatory (Kyoto University, Japan) and those obtained by 
the Solar Optical Telescope on board the {\it Hinode} satellite.}
}

%

\author{Yoichi \textsc{Takeda}}
\affil{National Astronomical Observatory, 2-21-1 Osawa, 
Mitaka, Tokyo 181-8588}
\email{takeda.yoichi@nao.ac.jp}
\and
\author{Satoru \textsc{UeNo}}
\affil{Kwasan and Hida Observatories, Kyoto University,
Kurabashira, Kamitakara, Takayama, Gifu 506-1314}
\email{ueno@kwasan.kyoto-u.ac.jp}

%


\maketitle

\begin{abstract}
In incorporating the effect of atmospheric turbulence in the broadening 
of spectral lines, the so-called radial-tangential macroturbulence (RTM) 
model has been widely used in the field of solar-type stars, which was
devised from an intuitive appearance of granular velocity field of the Sun.
Since this model assumes that turbulent motions are restricted to 
only radial and tangential directions, it has a special broadening 
function with notably narrow width due to the projection effect,
the validity of which has not yet been confirmed in practice.
With an aim to check whether this RTM model adequately represents the 
actual solar photospheric velocity field, we carried out an extensive 
study on the non-thermal velocity dispersion along the line-of-sight 
($V_{\rm los}$) by analyzing spectral lines at various points of 
the solar disk based on locally-averaged as well as high 
spatial-resolution spectra, and found the following results.    
First, the center-to-limb run of $V_{\rm los}$ derived from ground-based 
low-resolution spectra is simply monotonic with a slightly increasing 
tendency, which contradicts the specific trend (an appreciable peak 
at $\theta \sim 45^{\circ}$) predicted from RTM.  
Second, the $V_{\rm los}$ values derived from a large number of
spectra based on high-resolution space observation 
revealed to follow a nearly normal distribution, without any sign 
of peculiar distribution expected for the RTM case.
These two observational facts indicate that the actual solar velocity 
field is not such simply dichotomous as assumed in RTM, but
directionally more chaotic. 
We thus conclude that RTM is not an adequate model at least for 
solar-type stars, which would significantly overestimate the turbulent 
velocity dispersion by a factor of $\sim 2$. The classical Gaussian 
macroturbulence model should be more reasonable in this respect.
\end{abstract}

%


\section{Introduction}

The Doppler effect of atmospheric turbulence on the broadening of spectral 
lines plays a significant role in stellar spectroscopy. For example,
in spectroscopic determination of projected rotational velocities 
($v_{\rm e}\sin i$) of solar-type (FGK-type) stars, it is crucial to
properly eliminate the line-broadening component of turbulence origin, 
because it is comparable to (or predominant over) the rotational 
broadening in such generally slow rotators decelerated due to 
the magnetic braking mechanism. 

To make the problem easy and tractable, a very rough approximation 
has been adopted in traditional stellar spectroscopy, where turbulence 
in stellar atmospheres is divided into ``micro''-turbulence and 
``macro''-turbulence and separately treated, where the former 
(microscopic scale) is included in the Doppler with of the 
line-opacity profile (like thermal velocity) while the latter 
(macroscopic scale) acts as a global velocity distribution
function (like rotational broadening function) to be convolved 
with the intrinsic profile. Given that the extent of the latter 
($\gtsim 2$~km~s$^{-1}$) is known to be comparatively larger 
and more important than the former ($\ltsim 1$~km~s$^{-1}$) 
in solar-type dwarfs, the latter ``macroturbulence'' is 
the main issue in this context.

Regarding the velocity distribution function of macroturbulence, 
the so-called ``radial-tangential macroturbulence'' (hereinafter 
abbreviated as RTM) model has been widely used so far,
which was introduced by Gray (1975) for the first time
for analyzing the line-profiles of late-type stars.
That is, the appearance of solar granular velocity field 
(consisting of convective cells moving upward/downward and horizontal 
motions between the rising/falling cells) inspired Gray (1975) to
postulate that the velocity vectors are directionally restricted 
to being either along stellar radius or tangential to the surface 
while the speed of gas motion in each direction follows the random 
Gaussian distribution with a dispersion parameter $\zeta_{\rm RT}$. 
Since then, along with the efficient Fourier transform technique 
(e.g., Smith \& Gray 1976), Gray and his coinvestigators have extensively 
applied this RTM model to line-profile analyses of  F-, G-, and K-type 
stars in various evolutionary stages (e.g., to determine $v_{\rm e}\sin i$ 
by separating $\zeta_{\rm RT}$; see, Gray 1988, 2005 for more details
regarding the technical descriptions and accomplished results in this field). 

However, we feel some concern regarding the applicability of RTM to the case 
of solar-type dwarfs. Namely, according to Gray (1984; cf. section~V therein),
the value of $\zeta_{\rm RT}$ ($\sim 4$~km~s$^{-1}$) derived from the flux spectrum of 
the Sun-as-a-star based on RTM is appreciably larger than the non-thermal dispersion 
or typical granular velocities ($\sim$2--3~km~m$^{-1}$) directly estimated 
from spectroscopic observations of the resolved-disk Sun. What is the cause of 
this difference?

In order to clarify this situation, we refer to the work of Takeda (1995b; 
hereinafter referred to as Paper~I). In paper~I, an extensive profile-fitting 
analysis was carried out for many ($\sim 300$) blend-free lines of various 
strengths in the solar flux spectrum by using RTM model with an aim to 
investigate the relation between $\zeta_{\rm RT}$ and the mean-formation 
depth ($\overline{\log\tau}$),\footnote{
As described in subsection~5.1 of Paper~I, the mean formation depth
in the solar flux spectrum ($\overline{\log\tau}$) is defined as 
$\overline{\log\tau} \equiv \int R^{0}_{\lambda}\log\tau_{5000}(\tau_{\lambda} =2/3)d\lambda/
\int R^{0}_{\lambda} d\lambda$, where $R^{0}_{\lambda}$ is the line depth of the intrinsic 
profile expressed as $R^{0}_{\lambda} \equiv 1.- F^{0}_{\lambda}/F^{0}_{\rm cont}$.
Note that the notation $\overline{\log\tau}$ is used for the line-forming depth 
in the flux spectrum as in Paper~I, while that for the intensity spectrum
is denoted as $\langle \log\tau \rangle$ in this paper [cf. equation~(9)].}
 and the following results were derived
(cf. figure~2 in Paper~I):\\
--- $\zeta_{\rm RT}$ progressively increases 
with depth from $\sim 2.3$~km~s$^{-1}$ (at $\overline{\log\tau} \sim -2$) to
$\sim 3.8$~km~s$^{-1}$ (at $\overline{\log\tau} \sim -0.5$).\\
--- While this depth-dependence of $\zeta_{\rm RT}$ is qualitatively 
consistent as compared to the tendency of solar photospheric non-thermal 
velocity dispersion ($V^{\rm rad}$, $V^{\rm tan}$; see, e.g., figures~1--3 
in Gurtovenko 1975c or figure~1 in Canfield \& Beckers 1976), the former is 
systematically higher by $\sim 1$~km~s$^{-1}$ for unknown reason,\footnote{
Although it was speculated in Paper~I that the limb effect might cause some 
extra broadening of line width in the disk-integrated flux spectrum, such an 
effect (even if any exists) is quantitatively too small to account for this excess.} 
which again confirmed that $\zeta_{\rm RT}$ tends to be larger than 
the directly estimated velocity dispersion.

It is worth noting here that previous determinations of non-thermal 
velocity dispersion mentioned above were done under
the assumption of anisotropic Gaussian distribution (cf. equation~(4)
in subsection~2.3; i.e., near-random distribution of velocity vectors), 
which is markedly different from the basic assumption of RTM. 
As a trial, we repeated the same analysis as done in Paper~I 
(with a fixed $v_{\rm e}\sin i$ of 1.9~km~s$^{-1}$) but with 
the classical Gaussian macroturbulence (hereinafter referred to as GM) 
expressed by one dispersion parameter $\eta$ instead of RTM. 
The resulting $\eta$ values are plotted against $\overline{\log\tau}$ 
in figure~1a, where the $\zeta_{\rm RT}$ vs. $\overline{\log\tau}$ relation 
is also shown for comparison. It is manifest from this figure that 
$\eta$ is systematically smaller than $\zeta_{\rm RT}$ (the difference 
amounting to a factor of $\sim 2$) and more consistent with the
literature results of $V^{\rm rad}$ or $V^{\rm tan}$.

This $\zeta_{\rm RT}$ vs. $\eta$ discrepancy is reasonably interpreted as due 
to the difference between the characteristic widths of these two broadening functions.
In order to demonstrate this point, the broadening functions for RTM ($M_{1}$)
and GM ($M_{2}$) are graphically displayed in figures~1b and 1c, respectively.
Focusing on the zero-rotation case ($v_{\rm e}\sin i = 0$), we see 
that the half-width at half maximum (HWHM) for $M_{1}$ is 
HWHM$_{1} = 0.36 \zeta_{\rm RT}$ while that for $M_{2}$ is HWHM$_{2} = 0.83 \eta$, 
which yields $\zeta_{\rm RT}/\eta \sim 2.3 (\simeq 0.83/0.36)$ by equating these
two widths as HWHM$_{1}$ = HWHM$_{2}$. 
That is, since the width of RTM is narrower than that of GM (see also Fig.~17.5 of 
Gray 2005), the inequality of $\zeta_{\rm RT} > \eta$ generally holds 
regarding the solutions of $\zeta_{\rm RT}$ and $\eta$ required to reproduce 
the observed line width. This should be the reason why $\zeta_{\rm RT}$ is larger than
$\eta$ by a factor of $\sim 2$. 

Given that resulting solutions of macroturbulence are so significantly dependent 
upon the choice of broadening function, it is necessary to seriously consider 
which model represents the actual velocity field more adequately. 
Especially, we wonder whether RTM is ever based on a reasonable assumption,
because its broadening width is appreciably narrower than the extent of turbulent 
velocity dispersion, which stems from the extraordinary two-direction-confined 
characteristics (i.e., due to the projection effect; cf. figure~2).
Since the peculiarity of RTM lies in its specific angle-dependence, we would 
be able to give an answer to this question by studying the solar photospheric 
velocity dispersion at various points on the disk (i.e., from different view angles).

According to this motivation, we decided to challenge the task of verifying 
the validity of RTM by using the Sun as a testbench. Our approach is simple and 
straightforward in the sense that we carefully examine the widths of spectral 
lines from the disk center to the limb by making use of the profile-fitting
technique (as adopted in Paper~I), by which the widths of local broadening function 
can be efficiently determined while eliminating the effects of 
intrinsic and instrumental profiles. 
Regarding the observational data, we employed two data sets of local intensity 
spectra taken at a number of points on the solar disk: (i) spatially averaged 
spectra (over $\sim 50''$) to test whether the center-to-limb variation 
of the macrobroadening width predicted by RTM is observed, and (ii) spectra of 
fine spatial resolution (with sampling step of $\sim 0.''1$--$0.''3$) to examine 
the validity of the fundamental assumption on which RTM stands.

The remainder of this article is organized as follows:
We first explain the definitions of RTM and GM models in section~2, 
which forms the fundamental basis for the following sections.
Section~3 describes our method of analysis using the profile-fitting technique.  
In section~4, the behavior of velocity dispersion along the line-of-sight 
is investigated based on the low-resolution ground-based data and compared with 
the prediction from RTM. The analysis of high-resolution data from space observation 
is presented in section~5, where statistical properties of local velocity dispersion 
and radial velocity are discussed. The conclusions are summarized in section~6.
In addition, three special appendices are provided, where the influence of the choice
of microturbulence is checked (appendix~1), solar depth-dependent non-thermal 
velocity dispersions are derived to compare with the literature results (appendix~2), 
and the behavior of macroturbulence in solar-type stars is discussed (appendix~3).

\section{Definition of macroturbulence broadening function}

In this section, we briefly describe the basic definitions of representative 
macroturbulence broadening functions, which form the basis for the contents 
in later sections. 

\subsection{Line-profile modeling with macroturbulence}

In the approximation that the intrinsic specific intensity 
going to a direction angle $\theta$ [$I^{0}(v, \theta)$] is broadened by 
the local macroturbulence function [$\Theta(v, \theta)$],  
the emergent intensity profile [$I(v, \theta)$] is expressed as
\begin{equation}
I(v, \theta) = I^{0}(v, \theta) \otimes \Theta(v, \theta),
\end{equation}
where $\otimes$ means ``convolution.''

Similarly, when the intrinsic stellar flux [$F^{0}(v)$] is broadened by the 
integrated macroscopic line-broadening function [$M(v)$] (including the 
combined effects of macroturbulence and rotation), we may write the finally 
resulting flux profile [$F(v)$] as 
\begin{equation}
F(v) = F^{0}(v) \otimes M(v),
\end{equation} 
where an implicit assumption is
made that the continuum-normalized profile of $F^{0}$ does not vary over the 
stellar disk. 
Generally, $M(v)$ is derived by integrating $\Theta(v, \theta)$ over the disk, while 
appropriately Doppler-shifting (with the assumed $v_{\rm e}\sin i$) as well as 
multiplying by the limb-darkening factor (see Gray 1988 or Gray 2005 for more details).

\subsection{Radial--tangential macroturbulence}

Regarding the widely used radial-tangential macroturbulence (RTM) model (Gray 1975),
$\Theta(v, \theta)$ is defined as
\begin{eqnarray}
\Theta_{1}(v, \theta) = 
\frac{A_{\rm R}}{\pi^{1/2}\zeta_{\rm R}\cos\theta} \exp[-v^{2}/(\zeta_{\rm R}\cos\theta)^2]+ \nonumber \\
\frac{A_{\rm T}}{\pi^{1/2}\zeta_{\rm T}\sin\theta} \exp[-v^{2}/(\zeta_{\rm T}\sin\theta)^2],
\end{eqnarray}
though $A_{\rm R} = A_{\rm T}$ and $\zeta_{\rm R} = \zeta_{\rm T} (= \zeta_{\rm RT})$ are
usually assumed to represent the macroturbulence by only one parameter ($\zeta_{\rm RT}$). 
It should be remarked that this is essentially a two-component model, in the sense
that fraction $A_{\rm R}$ and fraction $A_{\rm T}$ of the stellar surface are covered with 
region R and region T (respectively) and that intrinsic intensity spectrum in each region 
is broadened by only either (i.e., not both) of the radial or tangential turbulent flow
(cf. figure~2 for a schematic description of this model). 
In practice, direct application of $\Theta_{1}$ to solar intensity spectrum is difficult, 
since it has an unusual profile especially near to the disk center ($\sin \theta \sim 0$) or 
to the limb ($\cos \theta \sim 0$). That is, as the broadening function is defined by 
the sum of a Gaussian profile (with a reasonable width) and a $\delta$-function-like profile 
(with very narrow width and very high peak), its width at half-maximum 
does not represent the real velocity dispersion any more (cf. figure~3a and figure~3a'). 
Accordingly, this model is used primarily in stellar application after the integration 
over the disk has been completed. 
Figure~1b shows the profiles of the disk-integrated macrobroadening function 
for this model $M_{1}(v; \zeta_{\rm RT}, v_{\rm e}\sin i)$, which were computed 
for various values of $v_{\rm e}\sin i/\zeta_{\rm RT}$ ratio by following
the procedure described in Gray (1988) (with the assumption of rigid rotation
and limb-darkening coefficient of $\epsilon = 0.6$).

\subsection{Anisotropic Gaussian macroturbulence}

Alternatively, we can consider a Gaussian macroturbulence with an anisotropic 
character with respect to the radial and tangential direction.
This is the case where the intrinsic intensity spectrum is broadened by 
gas of near-random motions (in terms of both speed and direction) following
the Gaussian velocity distribution of ellipsoidal anisotropy
(with dispersions of $\eta_{\rm R}$ and $\eta_{\rm T}$ in the radial and
tangential direction, respectively).
Then, the local broadening function is expressed by the convolution of two Gaussians as
\begin{eqnarray}
\Theta_{2}(v, \theta) \propto 
        \frac{ \exp[-v^{2}/(\eta_{\rm R}\cos\theta)^2] }{ \pi^{1/2}\eta_{\rm R}\cos\theta } 
\otimes \frac{ \exp[-v^{2}/(\eta_{\rm T}\sin\theta)^2] }{ \pi^{1/2}\eta_{\rm T}\sin\theta } \nonumber \\
\propto \exp\Bigl[-\frac{v^{2}}{(\eta_{\rm R}\cos\theta)^2+(\eta_{\rm T}\sin\theta)^2}\Bigr].
\end{eqnarray}
Actually, this is the traditional turbulence model which was used by solar physicists 
in 1960s--1970s to derive the radial and tangential components of non-thermal velocity 
dispersions (cf. section~1).
In the special case of $\eta_{\rm R} = \eta_{\rm T} (= \eta)$, equation~(4)
reduces to the simple isotropic Gaussian function
\begin{equation}
\Theta_{2}(v) \propto \exp[-(v/\eta)^2].
\end{equation}
By integration of $\Theta_{2}(v)$ over the disk, the integrated macrobroadening function 
$M_{2}(v; \eta, v_{\rm e}\sin i)$ can be obtained.
Figure~1c display the profiles of $M_{2}(v; \eta, v_{\rm e}\sin i)$ for 
different $v_{\rm e}\sin i/\eta$ ratios, which were numerically computed in the same manner 
as in $M_{1}$ (though $M_{2}$ in this case of angle-independent $\Theta_{2}$ can be expressed 
by a simple convolution of rotational broadening function and Gaussian function).

\section{Profile fitting for parameter determination}

Regarding the profile-fitting analysis (to be described in the following two sections), 
almost the same procedure as in Paper~I was adopted,
except that (i) specific intensity ($I$) emergent with angle $\theta$ 
corresponding to each observed point is relevant here (instead of angle-integrated 
flux) and (ii) Gaussian line-broadening function parameterized by $V_{\rm los}$ 
(velocity dispersion along the line of sight) was used for the kernel function:
\begin{equation}
K(v) \propto \exp[-(v/V_{\rm los})^2].
\end{equation} 
That is, the intensity profile $I(v,\theta)$ emergent to direction angle 
$\theta$ is expressed as
\begin{equation}
I(v,\theta) = I^{0}(v,\theta) \otimes K(v) \otimes P(v),
\end{equation}
where $P(v)$ is the instrumental profile. 
$I^{0}(v,\theta)$ is the intrinsic profile of outgoing specific intensity 
at the surface, which is written by the formal solution of radiative transfer as 
\begin{equation}
I^{0}(\lambda; \theta) = \int_{0}^{\infty} S_{\lambda}(t_{\lambda}) 
\exp(-t_{\lambda}/\cos\theta)d(t_{\lambda}/\cos\theta),
\end{equation}
where $S_{\lambda}$ is the source function and $t_{\lambda}$ ts the optical depth
in the vertical direction. 
Regarding the calculation of $I^{0}$, we adopted Kurucz's (1993) ATLAS9 solar 
photospheric model with a microturbulent velocity of $\xi$ = 0.5~km~s$^{-1}$ 
(see appendix~1 regarding the effect of changing this parameter) while assuming LTE.

Following Paper~I, we adopted the algorithm described in Takeda (1995a) 
to search for the best-fit theoretical profile, where the following three 
parameters were varied for this purpose: $\log\epsilon$ (elemental abundance), 
$V_{\rm los}$ (line-of-sight velocity dispersion), and  $\Delta\lambda_{\rm r}$ 
(wavelength shift) [which is equivalent to the radial velocity $v_{\rm r}$
($\equiv c \Delta\lambda_{\rm r}/\lambda$; c: velocity of light)]. 
 
After the solutions of these parameters have been converged, we computed 
the mean depth of line formation ($\langle \log \tau \rangle$) defined as follows:
\footnote{
Note that equation~(9) (see also footnote~1 for the flux case) 
is one of the various possibilities of defining line-forming depth.
Actually, photons of a given wavelength in a profile naturally 
emerge (not from a certain depth but) from a wide region 
according to the contribution function, for which several
definitions are proposed (see, e.g., Magain 1986).
Besides, there is no appointed procedure regarding how to 
represent the mean-formation depth of a line ``as a whole''
from the different forming depths for each of the wavelength points
within a profile. See, e.g., Gurtovenko and Sheminova (1997) 
for a review on the formation region of spectral lines.  
At any rate, the difference between our adopted definition and 
other ones is not so significant in the practical sense (cf. 
subsection~2.3 of Takeda 1992).
}
\begin{equation}
\langle \log \tau \rangle \equiv \frac{\int R^{0}_{\lambda}\log 
 \tau_{5000}(\tau_{\lambda} = \cos\theta) d\lambda} {\int R^{0}_{\lambda} d\lambda}
\end{equation}
where $\tau_{5000}$ is the continuum optical depth at 5000~$\rm\AA$,
$R^{0}_{\lambda}$ is the line depth of theoretical intrinsic profile 
(corresponding to the resulting solution of $\log\epsilon$) with respect to 
the continuum level ($R^{0}_{\lambda} \equiv 1 - I^{0}_{\lambda}/I^{0}_{\rm cont}$),
and integration is done over the line profile.
Besides, the local equivalent width ($w_{\lambda}^{i}$) could be evaluated
as a by-product by integrating $R^{0}_{\lambda}$ over the wavelength.

\section{Analysis of low spatial-resolution spectra}

\subsection{Expected angle-dependence of $V_{\rm los}$}

We are now ready to test the validity of RTM based on actual spectra
at various points of the solar disk. As RTM is a two-component model,
which is meaningful only in the combination of radial- and tangential-flow
parts, the observed spectra to be compared must be locally averaged
over a sufficient number of granular cells. 

Since $\Theta_{1}(v, \theta)$ has an extraordinary form (cf. subsection~2.2) 
and can not be directly incorporated in the analysis scheme described 
in section~3, we proceed with the following strategy:\\
--- Let us first assume that the RTM model exactly holds, which is characterized 
by two parameters ($\zeta_{\rm R}$ and $\zeta_{\rm T}$; 
while $A_{\rm R} = A_{\rm T}$ is assumed).\\
--- Then, the emergent intensity profile can be simulated by convolving
the intrinsic profile $R_{0}(v, \theta)$ with the RTM broadening function
$\Theta_{1}(v, \theta; \zeta_{\rm R}, \zeta_{\rm T})$ as
$R_{0} \otimes \Theta_{1}$.\\
--- Let us consider here how much $V_{\rm los}$ value would be obtained
if this RTM-broadened profile is analyzed by the Gaussian-based procedure described 
in section~3. This can be reasonably done by equating the HWHM (half-width at half-maximum) 
of $R_{0}(v, \theta) \otimes \Theta_{1}(v, \theta; \zeta_{\rm R}, \zeta_{\rm T})$
with that of $R_{0}(v, \theta) \otimes \exp[-(v/V_{\rm los})^2]$, by which we can express
$V_{\rm los}$ as a function of $\theta$ for any combination of ($\zeta_{\rm R}$, 
$\zeta_{\rm T}$).\\   
--- Regarding the intrinsic profile $R_{0}$, we adopted a Gaussian profile 
with $e$-folding half-width of 1.5~km~s$^{-1}$ (typical value for the thermal 
motion of Fe atom plus microturbulence in the solar photosphere). 

As examples, we display in figure~3 the profiles of $\Theta_{1}$ and 
$R_{0} \otimes \Theta_{1}$ computed for two cases ($\zeta_{\rm R} = \zeta_{\rm T}$ 
=  2 and 4~km~s$^{-1}$) at various angles ($\theta$).
The resulting $V_{\rm los}$ vs. $\theta$ relations for various combinations
of $(\zeta_{\rm R}, \zeta_{\rm T})$ are depicted (in solid lines) in figure~4, where 
the curves for the (anisotropic) Gaussian macroturbulence (GM) case are also shown 
(in dashed lines) for comparison
[$V_{\rm los}^{2} = (\eta_{\rm R} \cos\theta)^{2} +  (\eta_{\rm T} \sin\theta)^{2}$
holds in this GM case according to equation~(4)].

Several notable points are summarized below regarding the trends read from figure~4:\\
--- $V_{\rm los}$(GM) is almost the same order as $\sim \eta$, and thus
its dependence upon $\theta$ is nearly flat or monotonic.\\ 
--- On the contrary, $V_{\rm los}$(RTM) is significantly smaller 
than $\zeta$ [and $V_{\rm los}$(GM)] especially near to the disk center and near to the limb, 
resulting in a peak of $V_{\rm los}$(RTM) around $\theta \sim 45^{\circ}$. \\
--- This is due to the very characteristics of $\Theta_{1}$; i.e., its width 
tends to be considerably narrow without reflecting the turbulence dispersion 
(especially around $\theta \sim 0^{\circ}$ and $\theta \sim 90^{\circ}$) as clearly 
seen in figures~3a and 3a'.\\
--- Besides, why the inequality relation $V_{\rm los} < \zeta$ holds can also be 
understood from figure~2 (for the $\theta \sim 0^{\circ}$ case).\\  
--- Another important point is that this anomaly becomes particularly manifest 
when $\zeta$ outweighs $v_{\rm th}$ (1.5~km~s$^{-1}$), while less pronounced 
if $\zeta$ is comparable or smaller than $v_{\rm th}$.\\
We will make use of these characteristics to check the RTM model in subsection~4.4.

\subsection{Observational data of Hida/DST}

The ground-based observations were carried out on 2015 November 3--5
(JST) by using the 60~cm Domeless Solar Telescope (DST) with the
Horizontal Spectrograph at Hida Observatory of Kyoto University
(Nakai \& Hattori 1985).
The aspect angles of the solar rotation axis ($P$: position angle 
between the geographic north pole and the solar rotational north pole; 
$B_{0}$: heliographic latitude of the central point of the solar disk) 
in this period were $P = +24^{\circ}$ and $B_{0} = +4.^{\circ}1$.
Regarding the target positions on the solar disk, we selected 32 
points on the northern meridian line of the solar disk 
(from the disk center to 0.97~$R_{0}$ with a step of 
30$''\simeq$~0.03~$R_{0}$, where $R_{0}$ is the apparent radius 
of the solar disk) as depicted in figure~5a, at which the slit  
was aligned in the E--W direction.
Since the disk center and the nearest-limb point correspond to 
$\cos\theta =1$ and $\cos\theta = 0.24$ $(\equiv \sqrt{1-0.97^{2}})$ 
in this arrangement ($\theta$ is the emergent angle of the ray 
measured with respect to the normal to the surface), the angle
range of $0^{\circ}\le \theta \ltsim 76^{\circ}$ is covered by our data.

In the adopted setting of the spectrograph, 
our observation produced a solar spectrum covering of 153$''$ (spatial) 
and 24~$\rm\AA$ (wavelength) on the CCD detector with $1\times2$ binning 
(1600 pixels in the dispersion direction and 600 pixels in the spatial direction). 
We repeated the whole set (consecutive observations on 32 points
along the center-to-limb meridian) 30 times while changing the central 
wavelength, and finally obtained the spectra in the wavelength 
regions of 5190--5450~$\rm\AA$, 5650--5690~$\rm\AA$, 5830--5870~$\rm\AA$, 
and 6050--6310~$\rm\AA$ (about $\sim 600$~$\rm\AA$ in total).
Although most of our observations were done in quiet regions on
the solar disk, active regions may have affected in some of our data, 
since a notable spot passed through the meridian on November 4.
 
The data reduction was done by following the standard procedures
(dark subtraction, spectrum extraction, wavelength calibration,\footnote{
Since the wavelength vs. pixel relation was derived (not legitimately 
by using the comparison spectra but) based on $\sim$~20--50 solar lines in 
the disk-center spectrum for each region, any absolute wavelength calibration 
is not accomplished in our data.} and continuum normalization).  
The 1D spectrum was extracted by 
integrating over 200 pixels ($= 51''$; i.e., $\pm 100$ pixels 
centered on the target point) along the spatial direction.
Given that typical granule size is on the order of $\sim$~1$''$,
our spectrum corresponds to the spatial mean of each region including 
several tens of granular cells, by which the condition necessary for testing 
the RTM model is reasonably satisfied (cf. 1st paragraph of subsection~4.1). 
Finally, the effect of scattered light was corrected by following 
the procedure described in subsection~2.3 of Takeda and UeNo (2014), 
where the adopted value of $\alpha$ (scattered-light fraction) was 
0.10 ($\lambda < 5500$~$\rm\AA$) and 0.15 (($\lambda > 5500$~$\rm\AA$) 
according to our estimation.
Given that the main scope of this study is to measure the ``widths''
of spectral lines, this scattered-light correction does not 
have any essential influence, since it is a simple multiplication
of a factor to the line-depth profile (i.e., its similarity is
unaffected) as shown in equations (1) and (2) of Takeda and UeNo (2014).

The S/N ratio of the resulting spectrum (directly measured from 
statistical fluctuation in the narrow range of line-free continuum) 
turned out to be sufficiently high (typically $\sim$~500--1000).
The e-folding half width of the instrumental profile (assumed to be 
Gaussian as $\propto \exp (v/v_{\rm ip})^{2}$ in this study) 
was determined to be $v_{\rm ip} \simeq 1.3$~km~s$^{-1}$
by using the lamp + I$_{2}$ gas cell spectrum (cf. section~2 in 
Takeda \& Ueno 2012 for details), which corresponds to 
FWHM ($=2\sqrt{\ln 2} \; v_{\rm ip}$) $\simeq 2.2$~km~s$^{-1}$ and 
the spectrum resolving power of $\sim 140000 (\simeq c/{\rm FWHM})$ 
($c$: velocity of light).
Note that this $v_{\rm ip}$ of $\sim$~1.3~km~s$^{-1}$ is nearly 
the same order of the combined thermal+microturbulent velocity (e.g., 
$\sim 1.5$~km~s$^{-1}$ for the case of Fe atoms) and comparatively
smaller than the typical non-thermal velocity dispersion
($\sim$~2--3~km~s$^{-1}$; cf. appendix~2).

\subsection{Spectrum fitting and parameter determination}

Based on the same line list as used in Paper~I [originally 
taken from Meylan et al. (1993)], we selected a total of 86 lines  
corresponding to the wavelength ranges available in our data.
The basic data of these 86 lines (mostly Fe lines) are given in table~1, 
where we can see that lines of diversified strengths (equivalent widths 
from a few m$\rm\AA$ to $\sim 200$~m$\rm\AA$) are included. 
For convenience, we divided these 86 lines into 3 classes according
to the flux equivalent width ($W_{\lambda}^{f}$), as given in the table
(cf. the caption of the table).

We then applied the fitting technique described in section~3 to 
these lines. Quite a satisfactory convergence was successfully attained for 
most cases ($\sim 95$\%) of the total 2752 (= 32 points $\times$ 86 lines) 
trials, though solutions sometimes failed to converge or settled at
a poor fitting. 
 
As a demonstration, in figure~6 are compared the best-fit theoretical 
profiles with the observed profiles for three representative Fe~{\sc i} 
lines (weak, medium strength, strong) at three points (disk center,
intermediate angle of $\theta = 45^{\circ}$, near to the limb) on the disk.
Besides, the resulting values of $V_{\rm los}$, $\langle \log \tau \rangle$,
and $w_{\lambda}^{i}$ for these three lines at each point on the disk are 
plotted against $\theta$ in figure~7, where we can see that all these three
parameters tend to increase with an increase of $\theta$ and that stronger 
lines form at shallower layers.

We note in figure~7a that $V_{\rm los}$ shows appreciable fluctuations.
Actually, the solutions (especially of $V_{\rm los}$) appear to be rather sensitive to 
the local conditions as shown in figure~8, where the statistical distributions 
of $V_{\rm los}$ and $v_{\rm r}$ derived from a test analysis applied to 200 
disk-center spectra corresponding to each pixel (before spatially averaging) 
are shown. 

\subsection{Implication from the $V_{\rm los}$ vs. $\theta$ relation}

The resulting $V_{\rm los}$ solutions are plotted against $\theta$ in figure~9a (class-1 lines), 
figure~9b (class-2 lines), figure~9c (class-3 lines), and figure~9d (all lines).
As seen from the curve of growth depicted in figure~9e, each class
corresponds to weak unsaturated (class-1), moderately saturated (class-2),
and strongly saturated lines (class-3), respectively.
Among these three classes, the results derived from class-3 lines 
($W_{\lambda}^{f} \ge 100$~m$\rm\AA$) had better be viewed with some caution, 
because of a different sensitivity to a choice of microturbulence (cf. appendix~1).
We can see from figures~9a--c that the center-to-limb variation of $V_{\rm los}$ is 
characterized by a monotonic/gradual increase from $\sim 2$~km~s$^{-1}$ ($\theta \sim 0^{\circ}$)
to  $\sim$~2.5--3~km~s$^{-1}$ ($\theta \sim 80^{\circ}$). The trend that $V_{\rm los}$
tends to slightly decrease as the line becomes stronger may be attributed to its
depth-dependence (i.e., increasing with depth; cf. appendix~2),
since the line-forming depth becomes progressively shallower with the line strength 
(cf. figure~7b).   

Comparing the observed tendency of $V_{\rm los}$ (figures~9a--d) with the predicted 
trends for both RTM and GM (figure~4), we can draw a clear conclusion:
None of the $V_{\rm los}$(RTM) vs. $\theta$ relations matches the observed
center-to-limb variation (gradual increase), since the characteristic peak
at $\theta \sim 45^{\circ}$ expected for the RTM case is lacking. 
Let us recall that a significantly large  $\zeta_{\rm RT}$ value of $\sim$~3--4~km~s$^{-1}$
was derived from the analysis of solar flux spectrum (cf. section~1), which is evidently
larger than $v_{\rm th}$. Then, a prominent peak should be observed if the condition
assumed by RTM is really realized in the solar surface. Given the absence of such a
key trend, we can state that RTM is not a valid model for the solar atmospheric
velocity field. In contrast, the $V_{\rm los}$(GM) vs. $\theta$ curve predicted
for $(\eta_{\rm R}, \eta_{\rm T})$ = (2~km~s$^{-1}$, 3~km~s$^{-1}$) satisfactorily
reproduces the observed relation, which may indicate that the classical GM is 
a more reasonable and better representation in this respect.

\section{Analysis of high spatial-resolution spectra}

\subsection{Merit of studying well-resolved surface structures}

In order to ascertain the consequence of subsection~4.4 from an alternative 
point of view, we further carried out a similar analysis but using spectra 
of high spatial resolution acquired by satellite observations. 
Unlike the case of low-resolution spatially-averaged spectra 
studied in section~4, we can not employ these highly-resolved data 
for direct comparison with predictions from the RTM model, because
each spectrum reflects the gas motion of a local part    
smaller than the typical size of granules, to which the concept 
of RTM (meaningful only for spectra averaged over granular cells; cf. figure~2) 
is no more applicable. Instead, however, we can make use of such observational 
data of high spatial resolution to verify the fundamental assumption
on which the RTM model is based, since the velocity distribution (amplitude, 
direction) within a cell can be directly studied; e.g., whether the vectors of
turbulent motions are really coordinated in two orthogonal directions as 
assumed in RTM (cf.figure~2). This would make a decisive touchstone.

\subsection{Observational data of Hinode/SOT}

Regarding the spectra used for this purpose, we adopted the data obtained 
by the Solar Optical Telescope (SOT; Tsuneta et al. 2008) aboard the {\it Hinode}\footnote{
{\it Hinode} is a Japanese mission developed and launched by ISAS/JAXA, with NAOJ 
as domestic partner and NASA and STFC (UK) as international partners. 
It is operated by these agencies in co-operation with ESA and NSC (Norway).
} satellite (Kosugi et al. 2007). Since the Spectro-Polarimeter (SP; Lites et al. 2013)
in {\it Hinode}/SOT provides full calibrated Stokes $IQUV$ spectra of 
6301--6303~$\rm\AA$ region (comprising two Fe~{\sc i} lines at 6301.498 and 
6302.494~$\rm\AA$), we could use unpolarized $I$ spectra for our purpose. 
which are available as Level-1 data from the Hinode Data Center\footnote{
$\langle$http://darts.isas.jaxa.jp/solar/hinode/$\rangle$.
} or from the SolarSoft site\footnote{
$\langle$http://sot.lmsal.com/data/sot/level1d/$\rangle$}.

Having inspected the archived data, we decided to use the spectra obtained by 
normal-map mode observations of three quiet regions along the southern meridian 
on 2008 December 17 (the start time of each mapping was 05:43:35, 09:34:05, and 
10:34:05 in UT, respectively). These SP mapping observations were done by moving 
the (N--S aligned) slit of $0.''16$ width in E--W direction by $\sim 0.''1$, 
and field-of-view of in the slit direction is $129''$ (corresponds to 408 pixels 
on the detector), resulting in sampling steps of $\sim 0.''1$ ($x$ or E--W direction) 
and $\sim 0.''3$ ($y$ or N--S direction).
Although the total region covered by each mapping was $30''\times 129''$,
we used only the spectra within three $20'' \times 20''$ square regions centered
at ($0''$, $0''$),  ($0''$, $-700''$), and ($0''$, $-975''$) corresponding to
the disk center ($\theta \simeq 0^{\circ}$), the half-right-angle view point 
($\theta \simeq 45^{\circ}$), and the limb ($\theta \simeq 80^{\circ}$),
respectively. Figure~5b indicates the locations of these three regions,
for which the numbers of the resulting spectra were 8191, 8266, and 8253, respectively. 

\subsection{Statistical properties of $V_{\rm los}$ and $v_{\rm r}$}

As done in subsection~4.3, we applied the spectrum-fitting method 
(cf. section~3) to these spectra and successfully established
the solutions of $V_{\rm los}$ and $v_{\rm r}$, where only 
Fe~{\sc i}~6302.494 line (the slightly weaker one of the 
available two lines; also included in table~1) was used
for this analysis, and the instrumental profile was assumed
to be the Gaussian function with the $e$-folding half-width 
of 0.71~km~s$^{-1}$ corresponding to the spectrum resolving power 
of $R \simeq 6302.5/0.025 \simeq 252000$ (cf. figure~7 of 
Lites et al. 2013).
The typical signal-to-noise ratio of these SOT spectra is 
around $\sim 100$.
Some selected examples of fitted theoretical and 
observed spectra are displayed in figure~10.
The histograms for the resulting $V_{\rm los}$ and $v_{\rm r}$, 
$V_{\rm los}$ vs. $v_{\rm r}$ correlation, and the continuum brightness 
vs. $v_{\rm r}$ relation, are graphically shown figure~11.

Although several notable trends are observed in figure~11 (e.g., 
blue-shift tendency of brighter points at the disk-center which should 
be due to rising hot bubbles), we here confine ourselves only to 
the main purpose of this study; i.e., checking the validity of RTM.  

Let us focus on the disk-center results ($\theta \sim 0^{\circ}$) 
shown in the top row of figure~12. If the condition assumed in RTM
is really existent in the solar surface (cf. figure~2), observations 
of disk center (i.e., line-of-sight normal to the surface) with 
high spatial resolution should reveal almost comparable numbers 
of cases with turblent-broadened profiles (R) and those with 
unbroadened sharp profiles (T). Then, the following trends are expected:\\
--- The distribution function of $V_{\rm los}$ would show an extraordinary 
feature (e.g., an appreciable hump at low $V_{\rm los}$).\\
--- Since cells of horizontal flow (T) show no radial velocity,
an unusually prominent peak would exist in the distribution of 
$v_{\rm r}$ at $v_{\rm r} \sim 0$.\\
--- As a consequence, a considerable bias at (small $V_{\rm los}$,  
$v_{\rm r} \sim 0$) would be observed in the $V_{\rm los}$ vs. 
$v_{\rm r}$ plot.

However, none of these features is observed in the disk-center results 
in figure~11, where we can confirm that $V_{\rm los}$ as well as $
v_{\rm r}$ follow a statistically near-normal distribution without 
any such expected bias as mentioned above. 
These observational facts suggest that the velocity vectors of solar 
photospheric turbulence are not confined to only two (radial and 
tangential) directions but more chaotic with rather random orientations. 
Accordingly, we have reached a decision that the basic assumption of RTM does not
represent the actual solar photosphere, which means that the RTM model 
does not correctly describe the spectral line-broadening of solar-type stars.

\section{Concluding remark}

We carried out an extensive spectroscopic investigation on the non-thermal 
velocity dispersion along the line-of-sight by analyzing spectral lines 
at various points of the solar disk, in order to check 
whether the RTM model (which has been widely used for line-profile 
studies of solar-type stars) adequately represents the actual solar 
photospheric velocity field.
Applying the profile-fitting analysis to two sets of observational data: 
(spatially-averaged spectra from Hida/DST observations and very high 
spatial-resolution spectra from {\it Hinode}/SOT observations), 
we found the following results.    

First, the center-to-limb variation of $V_{\rm los}$ derived from 
low-resolution spectra turned out simply monotonic with a slightly increasing 
tendency. This apparently contradicts the characteristic trend (an appreciable 
peak at $\theta \sim 45^{\circ}$) expected from the RTM model.  
Second, the distributions of $V_{\rm los}$ and $v_{\rm r}$ values derived from 
spectra of very high spatial resolution revealed to show a nearly normal 
distribution, without any sign of anomalous distribution predicted from
the RTM model.

These observational facts suggest that the fundamental assumption of RTM 
is not compatible with the real atmospheric velocity field of the Sun, 
which can not be so simple (i.e, being confined only to radial and tangential 
directions) but should be directionally more chaotic. We thus conclude that 
RTM is not an adequate model at least for solar-type stars.

It is evident that RTM significantly overestimates the turbulent velocity 
dispersion in the solar photosphere, which should actually be $\sim 2$~km~s$^{-1}$ 
(disk center) and $\sim 2.5$~km~s$^{-1}$ (limb) as evidenced from the mean (or peak)
value of $V_{\rm los}$ derived from high-resolution data indicated in figure~11
(leftmost panels). Therefore, the fact that RTM yields $\zeta_{\rm RT} \sim$~3--4~km~s$^{-1}$ 
for the solar macroturbulence (cf. section~1) simply means that the width 
of RTM broadening function ($M_{1}$) is unreasonably too narrow. 
We therefore stress that, when using RTM for 
analyzing line profiles of solar-type stars, $\zeta_{\rm RT}$ should be 
regarded as nothing but a fudge parameter without any physical meaning.
If it were carelessly associated with discussion of physical processes
(e.g., in estimation of the turbulent energy budget or in comparison 
with the sonic velocity), erroneous results would come out.

On the other hand, the classical Gaussian macroturbulence model should be 
more reasonable and useful in this respect. Actually,
our application of GM to the analysis of solar flux spectrum resulted in 
$\eta \sim 2$~km~s$^{-1}$ (cf. figure~1b). Likewise, the GM-based conversion formula
[equation~(A1)] lead to $V^{\rm rad} \sim 2$~km~s$^{-1}$ (at $\log\tau \sim -1.5$) 
and $V^{\rm tan} \sim 2.5$~km~s$^{-1}$ (at $\log\tau \sim -2$) as the non-thermal 
dispersion in radial and tangential direction (cf. figure~13 in appendix~2), which 
are in fairly good agreement with the directly evaluated results based on 
high-resolution observations mentioned above (note that 
the mean formation depth for Fe~{\sc i} 6302.494 line is
$\log\tau \sim -1.5$ for the disk center and $\log\tau \sim -2$ for the limb;
cf. figure~7b). Accordingly, application of the simple GM would be 
more recommended, rather than the inadequate and complex RTM. 
(See appendix~3, where the trend of macroturbulence in FGK-type dwarfs 
is discussed in view of applying the GM model.) 

Finally, some comments may be due on the future prospect 
in this field. Regarding the modeling of turbulent velocity field 
in the atmosphere of the Sun or solar-type stars, we have to 
mention the recent remarkable progress in the simulations of 
3D time-dependent surface convection (see, e.g., Nordlund, 
Stein, \& Asplund 2009, and the references therein), which 
successfully reproduce the observed characteristics of spectral 
lines (e.g., Asplund et al. 2000; Pereira et al. 2013, for the 
solar case) without any ad-hoc turbulent-velocity parameters (such as 
micro- and macro-turbulence in the classical case) and thus by far 
superior to the traditional modeling. However, given the enormous 
computational burden of calculating such elaborate 3D models, 
the simple micro/macro-turbulence model is expected to remain 
still in wide use for practical analysis of stellar spectra. 
Therefore, it would be very helpful if the behaviors of classical
microturbulence as well as macroturbulence can be predicted or
understood based on the realistic 3D simulations.
For example, while main emphasis is placed on shift and asymmentry 
(bisector) of spectral lines in demonstrating the predictions
of 3D models in comparison with observations, less attention seems 
to be paid to the ``width'' of spectral lines. Can the trend of 
apparent turbulent dispersion derived in this study (i.e., 
tangential component being slightly larger than the radial component, 
an increasing tendency with depth) be reproduced by such 
state-of-the-art 3D hydrodynamical models? Further contributions 
of theoreticians in this light would be awaited.

\bigskip

This work was partly carried out on the Solar Data Analysis System 
operated by the Astronomy Data Center in cooperation with the Hinode 
Science Center of the National Astronomical Observatory of Japan.

\appendix

\section{Influence of microturbulence on the results}

Regarding the classical microturbulence,\footnote{
This is by definition 
the microscopic turbulent velocity dispersion, the characteristic scale of 
which is assumed to be much smaller than the photon mean-free-path. As such, 
it is formally included into the Doppler width of line-opacity profile 
in parallel with the velocity of thermal motion.
} which is necessary to 
compute the theoretical intrinsic profile to be convolved with the macroscopic 
velocity distribution function, we adopted 0.5~km~s$^{-1}$ throughout 
this study in order to maintain consistency with Paper~I.
This is the value obtained by analyzing the profiles of 15 stronger lines
($\xi = 0.51 \pm 0.15$~km~s$^{-1}$; cf. subsection~5.1 in Paper~I), 
which is in fairly good agreement with Gray's (1977) result of
$\xi = 0.5 \pm 0.1$~km~s$^{-1}$ based on the Fourier transform analysis 
of line profiles.

Meanwhile, the solar $\xi$ values based on the conventional way using 
equivalent widths of spectral lines (i.e., by the requirement
that the resulting abundances do not show any systematic dependence
on the line strength) published in various literature tend to be 
$\simeq 1$~km~s$^{-1}$ (see, e.g., subsection~3.2 in Takeda 1994); i.e., 
somewhat larger than the profile-based value of 0.5~km~s$^{-1}$ we adopted.  
This discordance was already remarked in subsection~6.1 of Paper~I
and further discussed by Takeda et al. (1996; cf. subsection~4.3 therein).

However, which solar microturbulence (high-scale or low-scale) to choose 
would not play any essential role in deriving the macroscopic velocity 
dispersions ($V_{\rm los}$, $\zeta_{\rm RT}$, and $\eta$) mentioned in 
this study, since the former ($\sim$~0.5--1~km~s$^{-1}$) is quantitatively 
insignificant as compared to the combination of the latter 
($\gtsim$~1--2~km~s$^{-1}$) and the thermal velocity ($\sim 1.3$~km~s$^{-1}$ 
for the case of Fe) (note that each velocity dispersion contributes to the
total Doppler velocity in the form of ``(root-)square-sum'').

In order to see whether this argument is justified, we carried out extra test 
calculations where all the analysis were redone by using $\xi = 1.0$~km~s$^{-1}$
instead of the fiducial value of 0.5~km~s$^{-1}$.
In figure~12 are compared the resulting values of $V_{\rm los}^{1.0}$, 
$\zeta_{\rm RT}^{1.0}$, and $\eta^{1.0}$ with those of the standard
results ($V_{\rm los}^{0.5}$, $\zeta_{\rm RT}^{0.5}$, and $\eta^{0.5}$). 

We can see an interesting trend regarding the resulting difference 
of $V_{\rm los}^{1.0} - V_{\rm los}^{0.5}$, $\ldots$ etc.
That is, as long as weaker-strength regime is concerned ($w^{i}_{\lambda}$ or 
$W^{f}_{\lambda}$ is $\ltsim 100$~m$\rm\AA$), the macroscopic velocity dispersions 
tend to slightly decrease as a consequence of using larger $\xi$, which is 
reasonably understandable by considering the contribution of each velocity 
component to the total width as mentioned above. 
On the other hand, for the case of stronger lines ($\gtsim$~100--150~m$\rm\AA$), 
this tendency is inversed, and $V_{\rm los}^{1.0}$, $\zeta_{\rm RT}^{1.0}$, 
and $\eta^{1.0}$ turn out {\it larger} than the corresponding $\xi = 0.5$~km~s$^{-1}$
results. This phenomenon is attributed to the effect of $\xi$ on the degree of 
saturation. That is, since a line may get desaturated (or saturation may be 
retarded) by increasing $\xi$ as known in the traditional curve-of-growth analysis, 
a line for a given equivalent width (e.g., 150~m$\rm\AA$) is strongly saturated 
(i.e., boxed shape with wider width at the core) for the $\xi=0.5$~km~s$^{-1}$ case 
but not so for $\xi = 1.0$~km~s$^{-1}$. In such circumstances, the former contributes 
larger broadening than the latter to the total line width, by which this
trend ($V_{\rm los}^{1.0} > V_{\rm los}^{0.5}$, $\cdots$ etc) may be explained. 

In any event, figure~12 shows that the changes are only 
$\pm \ltsim 0.5$~km~s$^{-1}$ for most cases ($\pm \ltsim$~10--20\% 
may be a better estimation, since the amount of variation appears to be proportional 
to the absolute values). We may regard these differences as comparatively insignificant, 
though we should keep in mind that the effect of changing $\xi$ is different for 
stronger saturated lines (those with $\gtsim$~100--150~m$\rm\AA$ forming around 
$\log\tau \sim$~-1.5 to $\sim -2$) from that for other weaker lines.

\section{Nature of non-thermal velocity dispersion in the solar photosphere}

According to the conclusion of this paper, it is a good approximation
to represent the non-thermal turbulent velocity field in the solar photosphere
by an anisotropic Gaussian distribution with dispersions of $V^{\rm rad}$ 
(radial direction) and $V^{\rm tan}$ (tangential direction). 
Therefore, we can safely use the classical relation 
(see, e.g., section~3 in Gurtovenko 1975c and the references therein):
\begin{equation}
V_{\rm los}^{2} = (V^{\rm rad} \cos\theta)^{2} +  (V^{\rm tan} \sin\theta)^{2}
\end{equation}
[note that $V^{\rm rad}$ and $V^{\rm tan}$ are equivalent to $\eta_{R}$ and $\eta_{T}$
in equation~(4)].
Let us examine the quantitative characteristics of $V^{\rm rad}$ and 
$V^{\rm tan}$ by using the $V_{\rm los}$ data derived in section~4 
for many lines at various points on the solar disk.

We regard $V_{\rm los}$ values near to the disk center at 
$1 \ge \cos \theta > 0.95$ ($0^{\circ} \le \theta < 17^{\circ}$) 
as practically equivalent to $V^{\rm rad}$, which are plotted against
$\langle \log \tau\rangle$ in figure~13a (red symbols). Although 
the dispersion is rather large, we could draw a mean $V^{\rm rad}(\tau)$ 
relation (with an extrapolation at $\log \tau \ltsim -2$) as depicted 
in the solid line connecting the points at 
($\log \tau$, $V^{\rm rad}$) = ($-2.5$, 1.6), ($-2.0$, 1.8), ($-1.5$, 2.0),
($-0.5$, 2.3), and (0.0, 1.9). 
Similarly, we assume those $V_{\rm los}$ values near to the limb at
 $0.3 > \cos \theta$ ($73^{\circ} < \theta$) almost equivalent to 
$V^{\rm tan}$, which are shown in figure~13b (blue symbols). 
In this case, however, the number of points at the important region of 
$-1 \ltsim \log\tau \ltsim 0$ is insufficient. Therefore, we added the 
data points by making use of the $V_{\rm los}$ values (only for class-1
lines) observed at $0.3 < \cos \theta < 0.95$ ($17^{\circ} <\theta < 73^{\circ}$),
which were converted to $V^{\rm tan}$ with the help of equation~(A1) and
the mean $V^{\rm rad}(\tau)$ relation derived above, as plotted 
in green filled circles in figure~13b. Then, eye-inspecting the combined
trend of these symbols, we derived the mean $V^{\rm tan}(\tau)$  
relation as depicted in the dashed line connecting
($\log \tau$, $V^{\rm tan}$) = ($-2.5$, 2.3), ($-2.0$, 2.7), ($-1.5$, 2.9),
($-1.0$, 3.0). ($-0.5$, 2.9), and (0.0, 2.7).

Such derived mean relations of $V^{\rm rad}(\tau)$ and $V^{\rm tan}(\tau)$
are shown together and compared with the literature results in figures~13c
and 13d, from which the following characteristics are summarized:\\
--- Roughly speaking, our results may be regarded as almost consistent with 
the relations derived by the previous studies in terms of the general trend
and a quantitative agreement is seen.\\
--- Especially, we could confirm that $V^{\rm tan}$ is systematically larger 
than $V^{\rm rad}$ by $\sim 1$~km~s$^{-1}$, which was already reported in 
various literature.\\   
--- However, our results are not necessarily compatible with the simple 
picture of monotonically increasing $V^{\rm rad}$ and $V^{\rm tan}$ with depth
suggested by many of the previous investigations. According to our mean curves, 
the depth-increasing tendency of $V^{\rm rad}$ and $V^{\rm tan}$ manifestly seen 
at high layer ($\log \tau \sim -2$) reaches the ceiling around
$\log \tau \sim -1$ ($V^{\rm tan}$) or $\log \tau \sim -0.5$ ($V^{\rm rad}$),
below which the velocity dispersion turns to slightly decrease with depth.\\
--- Consequently, the relations we derived for $V^{\rm rad}$ and $V^{\rm tan}$
show not so much a systematically large sensitivity to depth as a rather 
weak depth-dependence with a broad maximum around $\log\tau \sim -1$.

\section{Related topics regarding the macroturbulence of solar-type stars}

We concluded in this paper that RTM is not good but the classical GM 
is more preferable as the macroturbulence model for solar-type dwarfs.
In connection with this consequence, we briefly mention below (1) the practical 
procedure for application of GM, (2) the conversion between the two systems, 
and (3) the empirical relation for estimating the macroturbulence.

As an example of application of Gaussian macroturbulence, we refer to the 
work of Takeda and Tajitsu (2009), who studied the properties of three solar twins.
In their modeling, the total macrobroadening function, $f_{\rm M} (v)$, 
was assumed to be the convolution of three Gaussian component functions 
$f_{\alpha} \propto \exp [-(v/v_{\alpha})^{2}]$, 
where $\alpha$ is any of ``ip'' (instrumental profile), 
``rt'' (rotation), and ``mt'' (macroturbulence); i.e.,
\begin{equation}
v_{\rm M}^{2} = v_{\rm ip}^{2}+v_{\rm rt}^{2}+v_{\rm mt}^{2}
    \; \; \; (= v_{\rm ip}^{2} + v_{\rm r+m}^{2}),
\end{equation}
where $v_{\rm r+m}$ is the ``macroturbulence+rotation.''
They related these broadening parameters ($v_{\rm ip}$, $v_{\rm rt}$, 
and $v_{\rm mt}$) to (resolving power $R$, $v_{\rm e}\sin i$, $\zeta_{\rm RT}$) as
$v_{\rm ip} \simeq (c/R)/(2\sqrt{\ln 2})$ ($c$ is the speed of light), 
$v_{\rm rt} \simeq 0.94 v_{\rm e}\sin i$, 
and $v_{\rm mt} \simeq 0.42 \zeta_{\rm RT}$. which are based on
the requirement that FWHMs of the relevant broadening profile and the Gaussian 
profile should be the same.\footnote{See footnote~12 in subsection~4.2 of 
Takeda, Sato, and Murata (2008) for a more detailed description regarding 
the derivation of these conversion relations.} 

Here, the factor (0.42) for conversion between $v_{\rm mt}$ and $\zeta_{\rm RT}$ 
was estimated from the pure radial-tangential and pure Gaussian profiles 
(i.e., no-rotation case of $M_{1}$ and $M_{2}$ in figures~1b and 1c), 
which was used to estimate $v_{\rm rt}$ from the $\zeta_{\rm RT}$ results 
obtained in Paper~I. However, this factor appears to somewhat depend on 
the additional broadening due to rotation. For example, the results of 
$\zeta_{\rm RT}$ and $\eta$ ($\equiv v_{\rm mt}$) derived for the case of 
solar flux spectrum ($v_{\rm e}\sin i = 1.9$~km~s$^{-1}$) presented in figure~1a, 
it is slightly larger as $\eta/\zeta_{\rm RT} \sim 0.6$.
Therefore, it is not much meaningful to discuss the precise value of 
this ratio, for which we may only state as being around $\sim 0.5$.  

Several empirical relations of $v_{\rm mac}$ (macroturbulence) expressed in terms 
of atmospheric parameters (e.g., $T_{\rm eff}$) have been proposed by several 
authors; e.g., Gray (1984), Valenti and Fischer (2005), Bruntt et al. (2010), 
and Doyle et al. (2014). Several points should be remarked regarding their
practical applications:\\
--- First, it should be clearly recognized on which macroturbulence model
the relation in question is based. 
Among the four references mentioned above, only Bruntt et al. (2010) used GM, while 
RTM was employed in the other three. This must be the reason why only Bruntt et al.'s (2010) 
$v_{\rm mac}$ values are appreciably lower than the others (cf. figure~4 
of Doyle et al. 2014).\footnote{This discrepancy was already pointed out 
in Sect. 4.1.3 of Doyle et al. (2014), though they do not seem to be clearly 
aware that the macroturbulence model adopted by Bruntt et al. (2010) is 
Gaussian and thus different from the others.}\\ 
--- Second, if $v_{\rm mac}$ values derived from the empirical relation 
are to be used for $v_{\rm e}\sin i$ determination, it is desirable to use 
them unchanged as they are, while following the same line-broadening 
model as that adopted in the original literature.
Easily assuming a simple scaling relation to convert it to other system 
is not recommendable, which may cause some unwanted errors. \\
--- Nevertheless, regarding the $v_{\rm mac}$ vs. $T_{\rm eff}$ relation
derived from the lower envelope of macroturbulence plus rotational velocity 
distribution such as done by Valenti and Fischer (2005; cf. figure~3 
therein),\footnote{It should be noted that Valenti and Fischer's (2005)
Equation~(1), which is their proposed relation between $v_{\rm mac}$ 
and $T_{\rm eff}$, includes an apparent typo: The sign before 
the $T_{\rm eff}$-dependent term should be `+', instead of `$-$'. 
That is, $v_{\rm mac} = 3.98 + (T_{\rm eff}-5770)/650$,
where $v_{\rm mac}$ is in km~s$^{-1}$ and $T_{\rm eff}$ is in K.} 
transformation by applying $v_{\rm mt} \simeq 0.42 \zeta_{\rm RT}$ 
(derived from $M_{1}$ and $M_{2}$ for the no-rotation case as mentioned above) may not be 
a bad approximation, since the lower envelope corresponds to $v_{\rm e}\sin i = 0$.  

Bearing these points in mind, we also tried to examine how the macroturbulence 
($v_{\rm mt}$) in solar-type stars depends on $T_{\rm eff}$ based on our 
own data. In figure~14 are plotted the $v_{\rm r+m}$ values against $T_{\rm eff}$ (symbols), 
which were determined from the spectrum-fitting analysis of 6080--6089~$\rm\AA$ region 
while assuming the simple Gaussian modeling for the line-broadening functions expressed 
by equation~(A2). These results were originally derived by Takeda and Honda (2005; 
FGK-type dwarfs stars), Takeda et al. (2007; solar analogs), and Takeda et al. 
(2013; Hyades stars). The lower envelope of this $v_{\rm r+m}$ distribution 
at 6000~K~$\ge T_{\rm eff} \ge$~5000~K can be fitted by the following analytical 
relation for $v_{\rm mt}$ (solid line in the figure):
\begin{equation}
v_{\rm mt} = 6.087\times 10^{1} -2.352\times 10^{-2}T_{\rm eff} 
+ 2.311\times 10^{-6} T_{\rm eff}^{2},
\end{equation}
where $v_{\rm mt}$ is in km~s$^{-1}$ and $T_{\rm eff}$ is in K.
For comparison, the empirical relations published by Gray (1984), Valenti and Fischer (2005), 
Bruntt et al. (2010), and Doyle et al. (2014) are also depicted in this figure (dashed lines),
where only Bruntt et al.'s (2010) $v_{\rm mac}$ curve is shown unchanged (because of 
the same GM as $v_{\rm mt}$) while the $v_{\rm mac}$ values of other three are 
tentatively multiplied by a factor of 0.42 (because they are based on RTM). 
Roughly speaking, our result is more or less favorably compared with other relations,
as far as the lower temperature region (below the solar $T_{\rm eff}$) is concerned.
In particular,  those of Valenti and Fischer (2005) and Bruntt et al. (2010)
appear to show a reasonable consistency with our curve at $T_{\rm eff} \ltsim 5800$~K,

\clearpage
\onecolumn

\clearpage
\setcounter{table}{0}
\begin{table}[]
\caption{Data of adopted 86 spectral lines.}
\begin{center}
\scriptsize
\begin{tabular}
{cccrc c cccrc}
\hline \hline
Species & $\lambda$ & $\chi_{\rm low}$ & $W^{\rm f}_{\lambda}$ & class & &  Species & $\lambda $ & $\chi_{\rm low}$  & $W^{\rm f}_{\lambda}$ & class \\
   & ($\rm\AA$) & (eV) & (m$\rm\AA$) &  &  &  & ($\rm\AA$) & (eV) & (m$\rm\AA$) & \\
\hline
Fe~{\sc i}  & 5197.929 &  4.301 &   41.6 &   1 &  & Fe~{\sc i}  &  5651.470 &  4.473 &   22.1 &   1 \\
Fe~{\sc i}  & 5198.711 &  2.223 &  100.0 &   3 &  & Fe~{\sc i}  &  5652.320 &  4.260 &   32.8 &   1 \\
Fe~{\sc i}  & 5206.801 &  4.283 &   12.1 &   1 &  & Fe~{\sc i}  &  5653.889 &  4.386 &   44.3 &   1 \\
Fe~{\sc i}  & 5223.187 &  3.635 &   33.3 &   1 &  & Fe~{\sc i}  &  5672.267 &  4.584 &    3.1 &   1 \\
Fe~{\sc i}  & 5225.525 &  0.110 &   74.7 &   2 &  & Fe~{\sc i}  &  5679.025 &  4.652 &   72.0 &   2 \\
Fe~{\sc ii} & 5234.625 &  3.221 &   96.9 &   2 &  & Fe~{\sc i}  &  5680.241 &  4.186 &   16.1 &   1 \\
Fe~{\sc i}  & 5242.491 &  3.634 &   94.3 &   2 &  & Fe~{\sc i}  &  6065.482 &  2.608 &  125.0 &   3 \\
Fe~{\sc i}  & 5247.049 &  0.087 &   67.2 &   2 &  & Fe~{\sc i}  &  6082.708 &  2.223 &   38.3 &   1 \\
Fe~{\sc i}  & 5253.023 &  2.279 &   21.8 &   1 &  & Fe~{\sc ii} &  6084.111 &  3.199 &   22.3 &   1 \\
Fe~{\sc i}  & 5279.654 &  3.301 &    8.2 &   1 &  & Fe~{\sc i}  &  6093.666 &  4.607 &   34.3 &   1 \\
Fe~{\sc i}  & 5285.118 &  4.434 &   32.8 &   1 &  & Fe~{\sc i}  &  6094.364 &  4.652 &   23.1 &   1 \\
Fe~{\sc i}  & 5288.528 &  3.695 &   65.2 &   2 &  & Fe~{\sc i}  &  6096.662 &  3.984 &   43.0 &   1 \\
Fe~{\sc i}  & 5300.412 &  4.593 &    8.6 &   1 &  & Ca~{\sc i}  &  6102.723 &  1.879 &  144.0 &   3 \\
Fe~{\sc i}  & 5301.327 &  4.386 &    4.3 &   1 &  & Fe~{\sc i}  &  6105.152 &  4.548 &   14.2 &   1 \\
Fe~{\sc i}  & 5308.707 &  4.256 &    9.8 &   1 &  & Fe~{\sc i}  &  6127.909 &  4.143 &   54.1 &   2 \\
Fe~{\sc i}  & 5320.039 &  3.642 &   23.5 &   1 &  & Fe~{\sc i}  &  6136.615 &  2.453 &  143.0 &   3 \\
Fe~{\sc i}  & 5321.109 &  4.434 &   47.7 &   1 &  & Fe~{\sc i}  &  6137.694 &  2.588 &  140.0 &   3 \\
Fe~{\sc ii} & 5325.553 &  3.221 &   49.7 &   1 &  & Fe~{\sc ii} &  6141.033 &  3.230 &    2.7 &   1 \\
Cr~{\sc i}  & 5348.312 &  1.004 &  105.0 &   3 &  & Fe~{\sc ii} &  6149.258 &  3.889 &   39.6 &   1 \\
Fe~{\sc i}  & 5364.858 &  4.446 &  138.0 &   3 &  & Fe~{\sc i}  &  6151.617 &  2.176 &   53.9 &   2 \\
Fe~{\sc i}  & 5365.396 &  3.573 &   85.0 &   2 &  & Fe~{\sc i}  &  6165.361 &  4.143 &   49.2 &   1 \\
Fe~{\sc i}  & 5367.479 &  4.415 &  159.0 &   3 &  & Ca~{\sc i}  &  6169.563 &  2.526 &  120.0 &   3 \\
Fe~{\sc i}  & 5376.826 &  4.294 &   18.0 &   1 &  & Fe~{\sc i}  &  6173.341 &  2.223 &   74.3 &   2 \\
Fe~{\sc i}  & 5379.574 &  3.695 &   67.2 &   2 &  & Fe~{\sc i}  &  6180.203 &  2.727 &   59.3 &   2 \\
Fe~{\sc i}  & 5385.579 &  3.695 &    5.8 &   1 &  & Fe~{\sc i}  &  6187.398 &  2.832 &    5.5 &   1 \\
Fe~{\sc i}  & 5389.479 &  4.415 &   97.6 &   2 &  & Fe~{\sc i}  &  6187.987 &  3.943 &   55.0 &   2 \\
Fe~{\sc i}  & 5395.215 &  4.446 &   26.3 &   1 &  & Fe~{\sc i}  &  6191.558 &  2.433 &  136.0 &   3 \\
Fe~{\sc i}  & 5398.277 &  4.446 &   84.8 &   2 &  & Fe~{\sc i}  &  6199.507 &  2.559 &    6.0 &   1 \\
Fe~{\sc i}  & 5401.264 &  4.320 &   29.5 &   1 &  & Fe~{\sc i}  &  6200.314 &  2.608 &   78.6 &   2 \\
Fe~{\sc i}  & 5406.770 &  4.371 &   42.7 &   1 &  & Fe~{\sc i}  &  6213.429 &  2.223 &   83.6 &   2 \\
Fe~{\sc i}  & 5409.133 &  4.371 &   63.1 &   2 &  & Fe~{\sc i}  &  6220.776 &  3.882 &   23.1 &   1 \\
Fe~{\sc i}  & 5412.798 &  4.434 &   25.2 &   1 &  & Fe~{\sc i}  &  6221.670 &  0.859 &    2.5 &   1 \\
Fe~{\sc ii} & 5414.073 &  3.221 &   34.7 &   1 &  & Fe~{\sc i}  &  6226.730 &  3.883 &   33.9 &   1 \\
Fe~{\sc i}  & 5415.192 &  4.386 &  199.0 &   3 &  & Fe~{\sc i}  &  6229.225 &  2.845 &   39.9 &   1 \\
Fe~{\sc i}  & 5417.039 &  4.415 &   39.9 &   1 &  & Fe~{\sc i}  &  6232.639 &  3.654 &   98.4 &   2 \\
Fe~{\sc ii} & 5425.257 &  3.199 &   50.1 &   2 &  & Fe~{\sc ii} &  6239.953 &  3.889 &   16.2 &   1 \\
Fe~{\sc i}  & 5436.297 &  4.386 &   45.2 &   1 &  & Fe~{\sc i}  &  6240.645 &  2.223 &   52.8 &   2 \\
Fe~{\sc i}  & 5436.587 &  2.279 &   50.1 &   2 &  & Fe~{\sc i}  &  6246.317 &  3.602 &  133.0 &   3 \\
Fe~{\sc i}  & 5441.354 &  4.312 &   37.0 &   1 &  & Fe~{\sc ii} &  6247.557 &  3.892 &   61.9 &   2 \\
Fe~{\sc i}  & 5443.409 &  4.103 &    5.2 &   1 &  & Fe~{\sc i}  &  6252.554 &  2.404 &  128.0 &   3 \\
Fe~{\sc i}  & 5445.042 &  4.386 &  130.0 &   3 &  & Fe~{\sc ii} &  6269.967 &  3.245 &    7.6 &   1 \\
Fe~{\sc i}  & 5650.020 &  5.099 &   41.8 &   1 &  & Fe~{\sc i}  &  6297.792 &  2.223 &   79.3 &   2 \\
Fe~{\sc i}  & 5650.704 &  5.085 &   47.8 &   1 &  & Fe~{\sc i}  &  6302.494 &  3.686 &   99.7 &   2 \\
\hline
\end{tabular}
\end{center}
Regarding the atomic data given here, the (air) wavelength and the lower excitation potential
were adopted from Kurucz and Bell's (1995) compilation. The solar flux equivalent widths
($W_{\lambda}^{\rm f}$; corresponding to the spectrum of the Sun-as-a star) were taken 
from table~1 and table~2 of Paper~I (mostly taken from Meylan et al. 1993), based on which 
the line-strength class was assigned to each line ($W_{\lambda}^{\rm f} < 50\;{\rm m\AA} \cdots$ 
class~1; $50\;{\rm m\AA} \le W_{\lambda}^{\rm f} < 100\;{\rm m\AA} \cdots$ class~2; 
$100\;{\rm m\AA} \le W_{\lambda}^{\rm f} \cdots$ class~3).
\end{table}

\clearpage

\begin{figure}
  \begin{center}
    \FigureFile(130mm,130mm){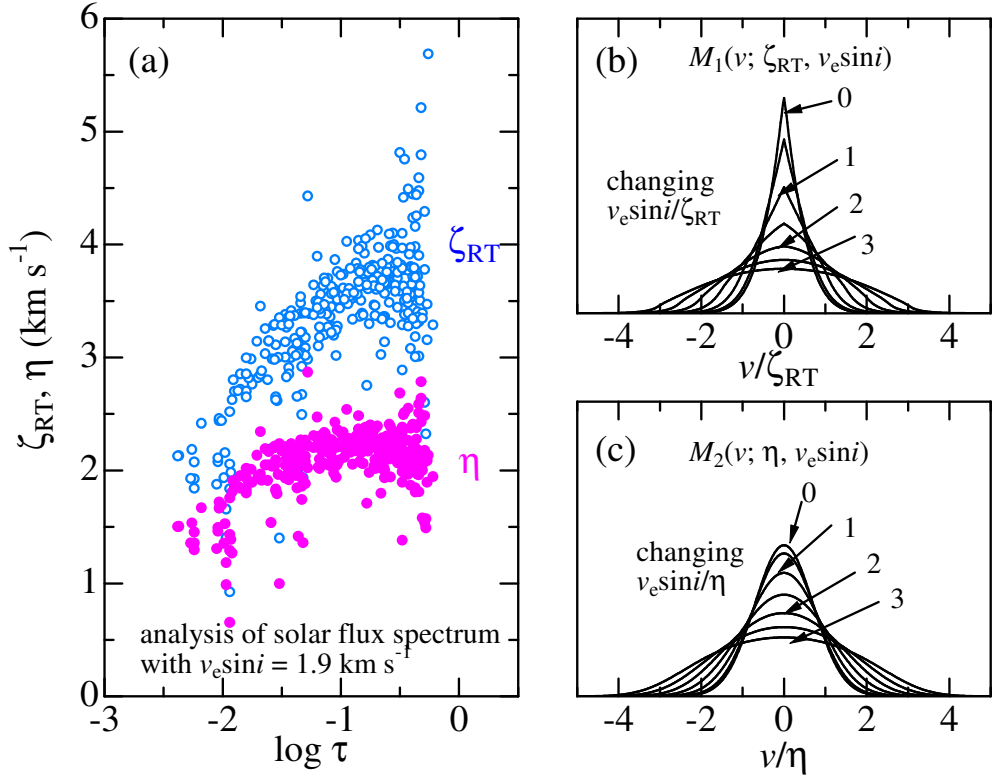}
  \end{center}
\caption{
(a) Results of radial--tangential macroturbulence ($\zeta_{\rm RT}$; open circles) 
and Gaussian macroturbulence ($\eta$; filled circles) plotted against 
$\langle \log \tau \rangle$, which were determined from our analysis of Kurucz et al.'s 
(1984) solar flux spectrum by two different macrobroadening functions ($M_{1}$ and $M_{2}$). 
(b) Radial-tangential macroturbulence plus rotational broadening function 
$M_{1}(v; \zeta_{\rm RT}, v_{\rm e}\sin i)$
(for seven $v_{\rm e}\sin i/\zeta_{\rm RT}$ values from 0.0 to 3.0 with an increment of 0.5) 
plotted against $v/\zeta_{\rm RT}$.
(c) Gaussian macroturbulence plus rotational broadening function 
$M_{2}(v; \eta, v_{\rm e}\sin i)$ (for seven $v_{\rm e}\sin i/\eta$ values) 
plotted against $v/\eta$.
}
\end{figure}

\begin{figure}
  \begin{center}
    \FigureFile(120mm,120mm){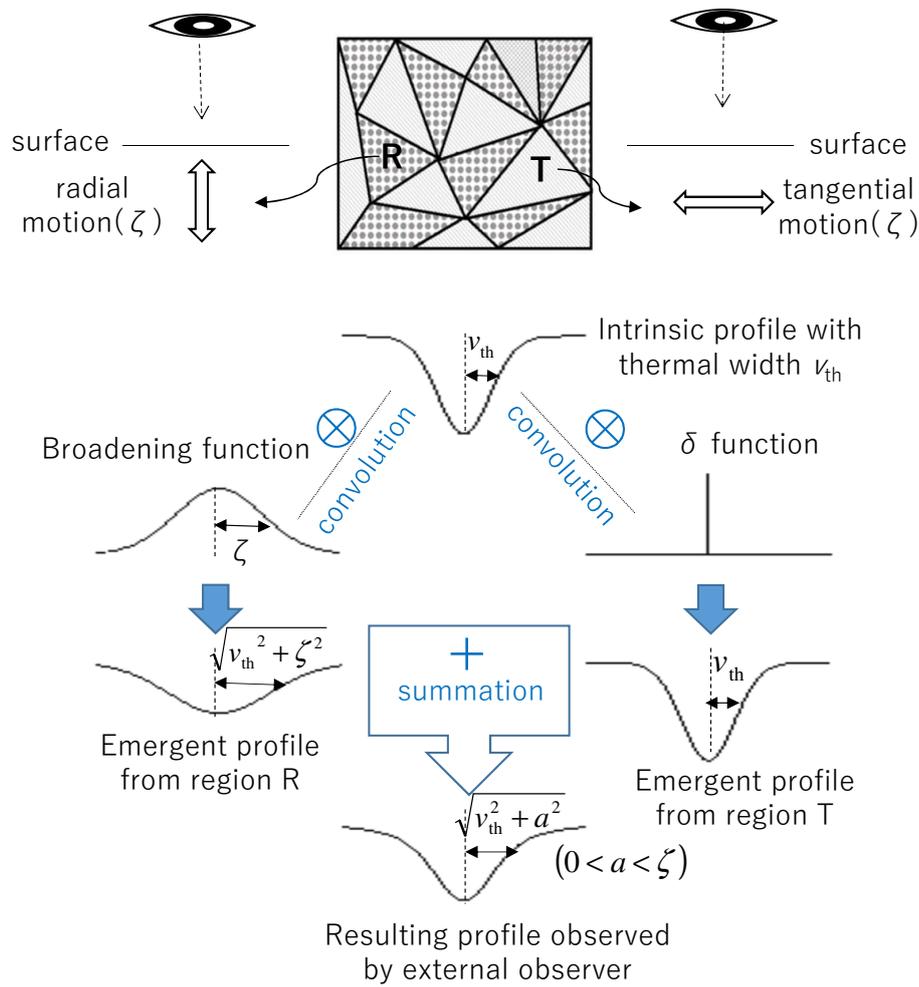}
  \end{center}
\caption{
Schematic description of the concept of radial--tangential macroturbulence
for the specific case of disk-center observation (the line-of-sight
of the observer is perpendicular to the solar surface),
which explains why the width of the broadening function is considerably
smaller than the actual velocity dispersion.
}
\end{figure}

\begin{figure}
  \begin{center}
    \FigureFile(100mm,160mm){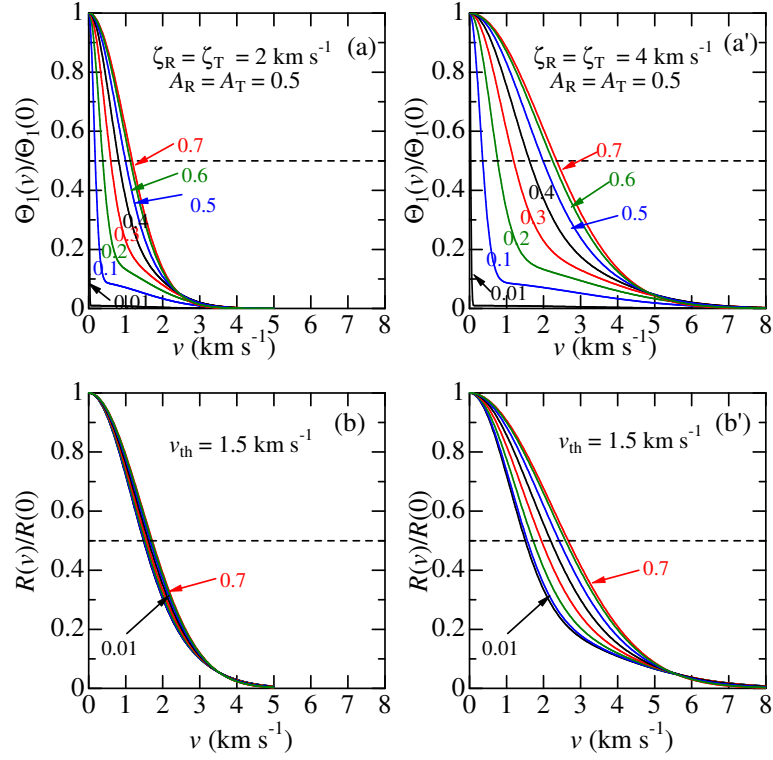}
  \end{center}
\caption{
(a) Local distribution function of radial-tangential macroturbulence [$\Theta_{1}(v)$]
expressed by equation~(3) corresponding to the parameters of
$\zeta_{\rm R} = \zeta_{\rm T} = 2$~km~s$^{-1}$ and $A_{\rm R} = A_{\rm T}$ = 0.5, 
computed for eight $\sin \theta$ values of 0.01, 0.1, 0.2, 0.3, 0.4, 0.5, 0.6, and 0.7.
(Note that only the results for $0^{\circ} < \theta < 45^{\circ}$ are given here,
since those for $45^{\circ} < \theta < 90^{\circ}$ can be derived 
by making use of the symmetric property in this case.)
The curves are normalized at the line-center ($v=0$), where the values of $\Theta_{1}(0)$
are 14.2, 1.55, 0.849, 0.618, 0.507, 0.445, 0.411, and 0.399, respectively.

(b) Line-depth profiles [$R(v)$; normalized at line-center] simulated by convolving 
$\Theta_{1}(v)$ with a Gaussian intrinsic profile having an $e$-folding half-width of 
$v_{\rm th} = 1.5$~km~s$^{-1}$. Panels (a') and (b') are almost the same as (a) and (b) 
but for $\zeta_{\rm R} = \zeta_{\rm T} = 4$~km~s$^{-1}$.
}
\end{figure}

\begin{figure}
  \begin{center}
    \FigureFile(150mm,150mm){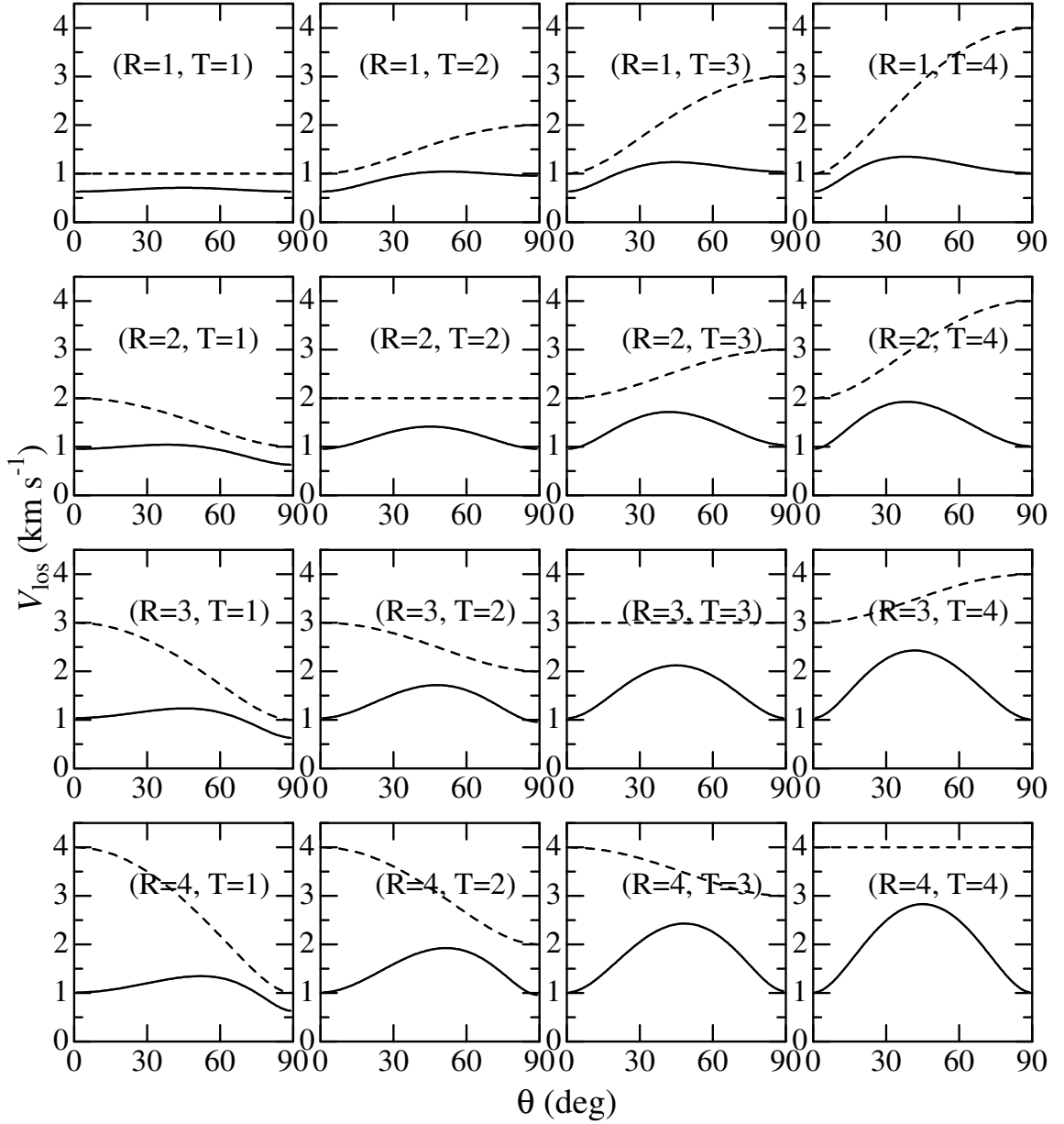}
  \end{center}
\caption{
Run of the theoretically expected values of $V_{\rm los}$ with $\theta$,
which were computed for different combinations of radial and tangential 
components of the assumed turbulence models (along with the Gaussian 
intrinsic thermal profile with an $e$-folding half-width of 
1.5~km~s$^{-1}$). Solid lines correspond to 
the case of radial--tangential macroturbulence, while dashed lines are 
the cases of anisotropic Gaussian macroturbulence. For example, 
in the panel indicated as (R=2, T=3), the solid line
shows the $\theta$-dependence of $V_{\rm los}$ computed for 
($\zeta_{\rm R} = 2$~km~s$^{-1}$, $\zeta_{\rm T} = 3$~km~s$^{-1}$), 
while the dashed line is for the result corresponding to 
($\eta_{\rm R} = 2$~km~s$^{-1}$, $\eta_{\rm T} = 3$~km~s$^{-1}$).
}
\end{figure}

\begin{figure}
  \begin{center}
    \FigureFile(90mm,90mm){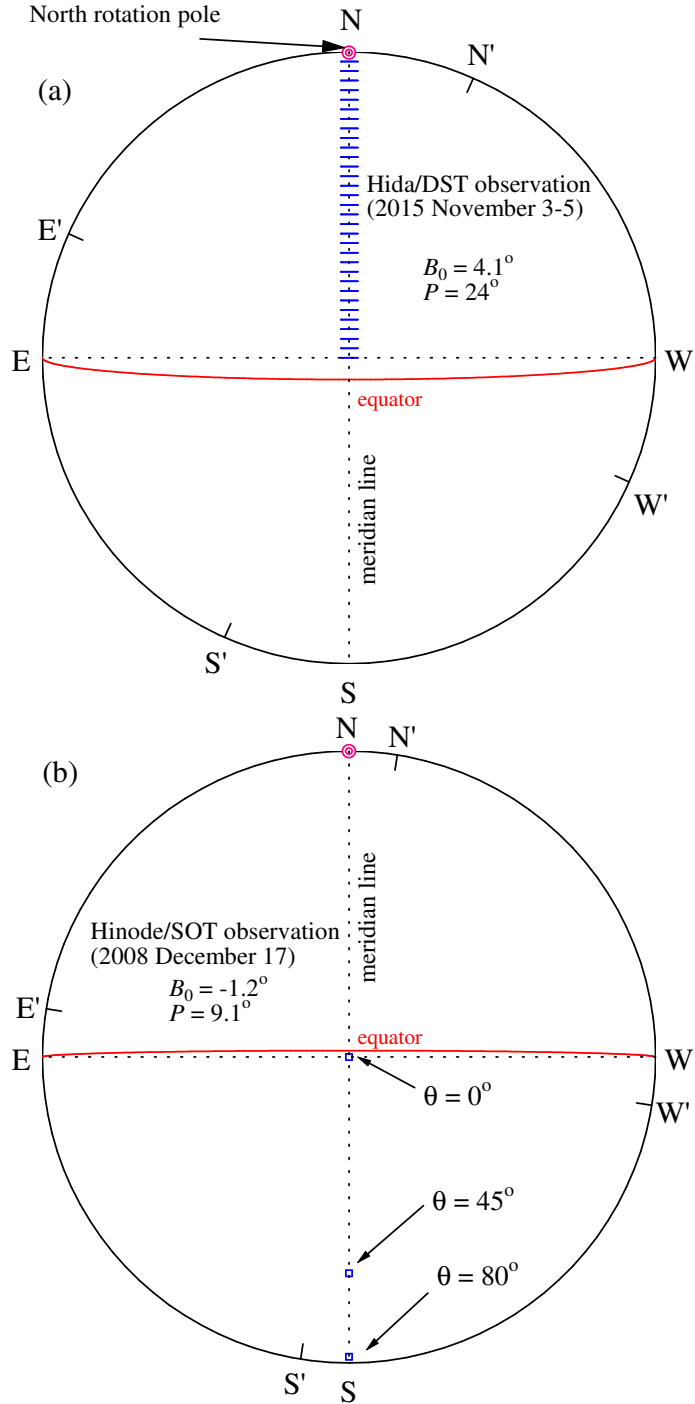}
  \end{center}
\caption{
Graphical description of the observed points on the solar disk,
at which the spectral data obtained in this study were taken.
The upper figure (a) corresponds to the Hida/DST observation on
2015 November 3--5 (32 points on the northern meridian from the disk center 
to $0.97 R_{\odot}$ with a step of $30'' \simeq 0,03R_{\odot}$, while
spatially averaged over $51''$ along the E--W direction).
The lower figure (b) is for the {\it Hinode}/SOT observation
on 2008 December 17 (three squares of $20''\times20''$ on the 
southern meridian at $\theta = 0^{\circ}$, $45^{\circ}$, and $80^{\circ}$).  
While N, S, E, and W are the directions in reference to the Sun (based on solar rotation), 
those in the equatorial coordinate system on the celestial sphere (defined by 
the rotation of Earth) are also denoted as N$'$, S$'$, E$'$, and W$'$.
}
\end{figure}

\begin{figure}
  \begin{center}
    \FigureFile(140mm,160mm){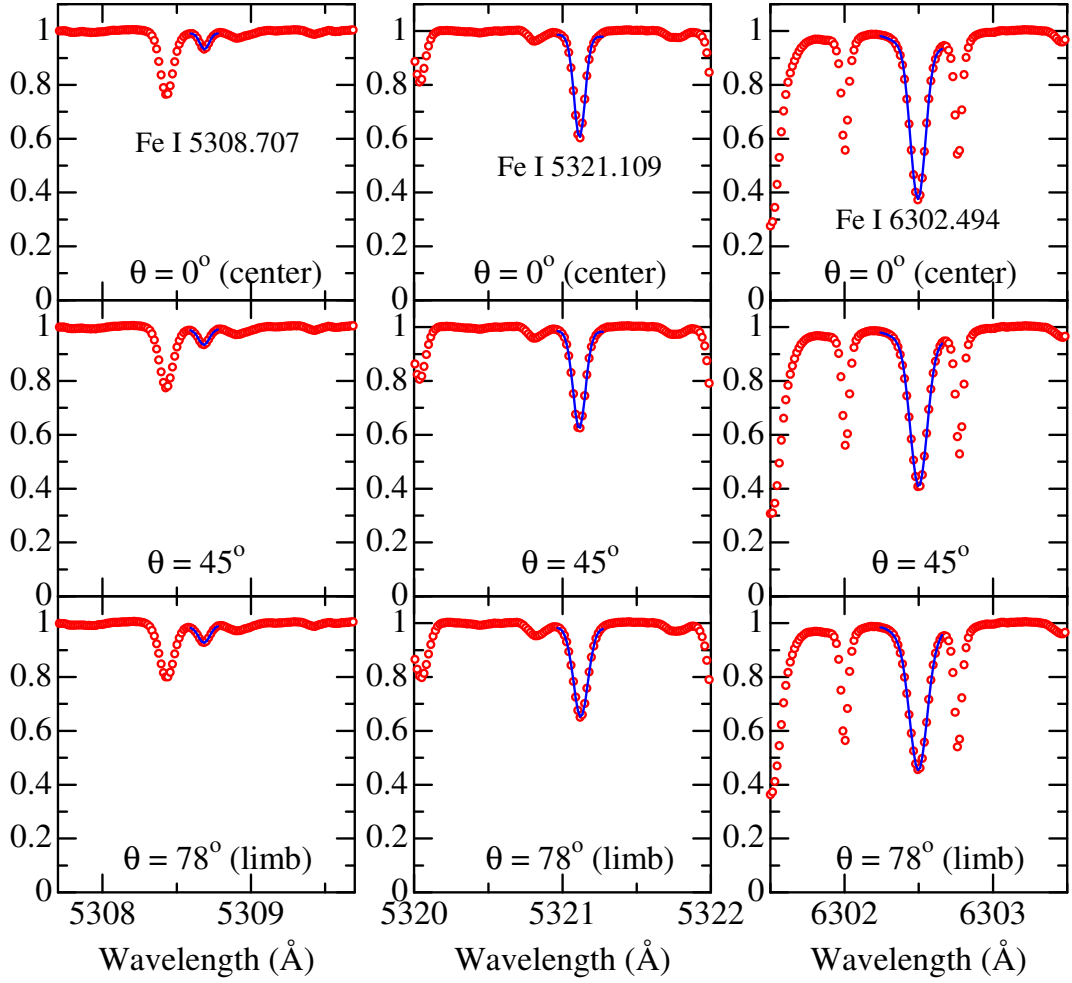}
  \end{center}
\caption{
Spectrum-fitting examples of Hida/DST data analysis.
Shown here are the selected representative cases of
weak (Fe~{\sc i} 5308.707; left panels), 
medium-strength (Fe~{\sc i} 5321.109; center panels),
and rather strong (Fe~{\sc i} 6302.494; right panels) lines. 
The panels in upper, middle, and lower row correspond to
disk center ($\theta = 0^{\circ}$), intermediate point 
($\theta = 45^{\circ}$), and near-limb ($\theta = 78^{\circ}$), respectively. 
The observed spectra are shown in red circles while the best-fit
theoretical spectra are depicted in blue lines.
}
\end{figure}

\begin{figure}
  \begin{center}
    \FigureFile(70mm,100mm){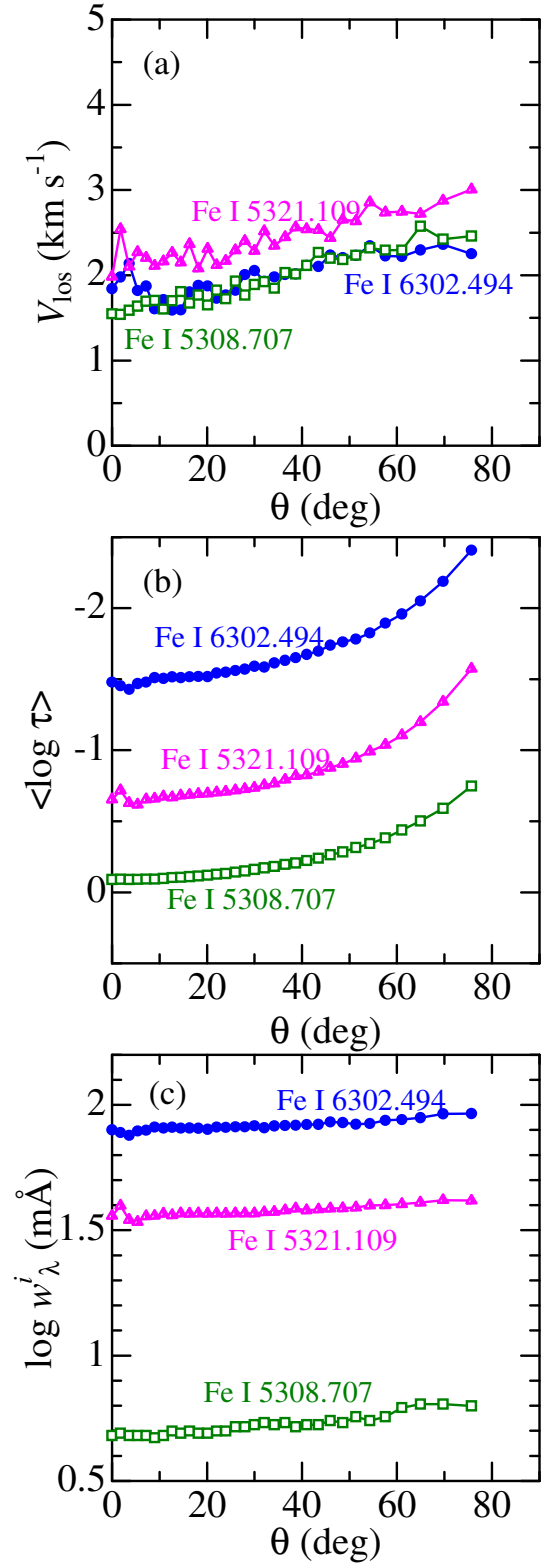}
  \end{center}
\caption{
Panels (a), (b), and (c) show the plots of $V_{\rm los}$ (line-of-sight 
velocity dispersion), $\langle \log \tau \rangle$ (mean depth of line formation), 
and $w_{\lambda}^{\i}$ (equivalent width) against $\theta$ (direction angle), 
respectively, which were derived from the analysis of Hida/DST data 
for three representative lines (the same lines as in figure~6). 
}
\end{figure}

\begin{figure}
  \begin{center}
    \FigureFile(150mm,150mm){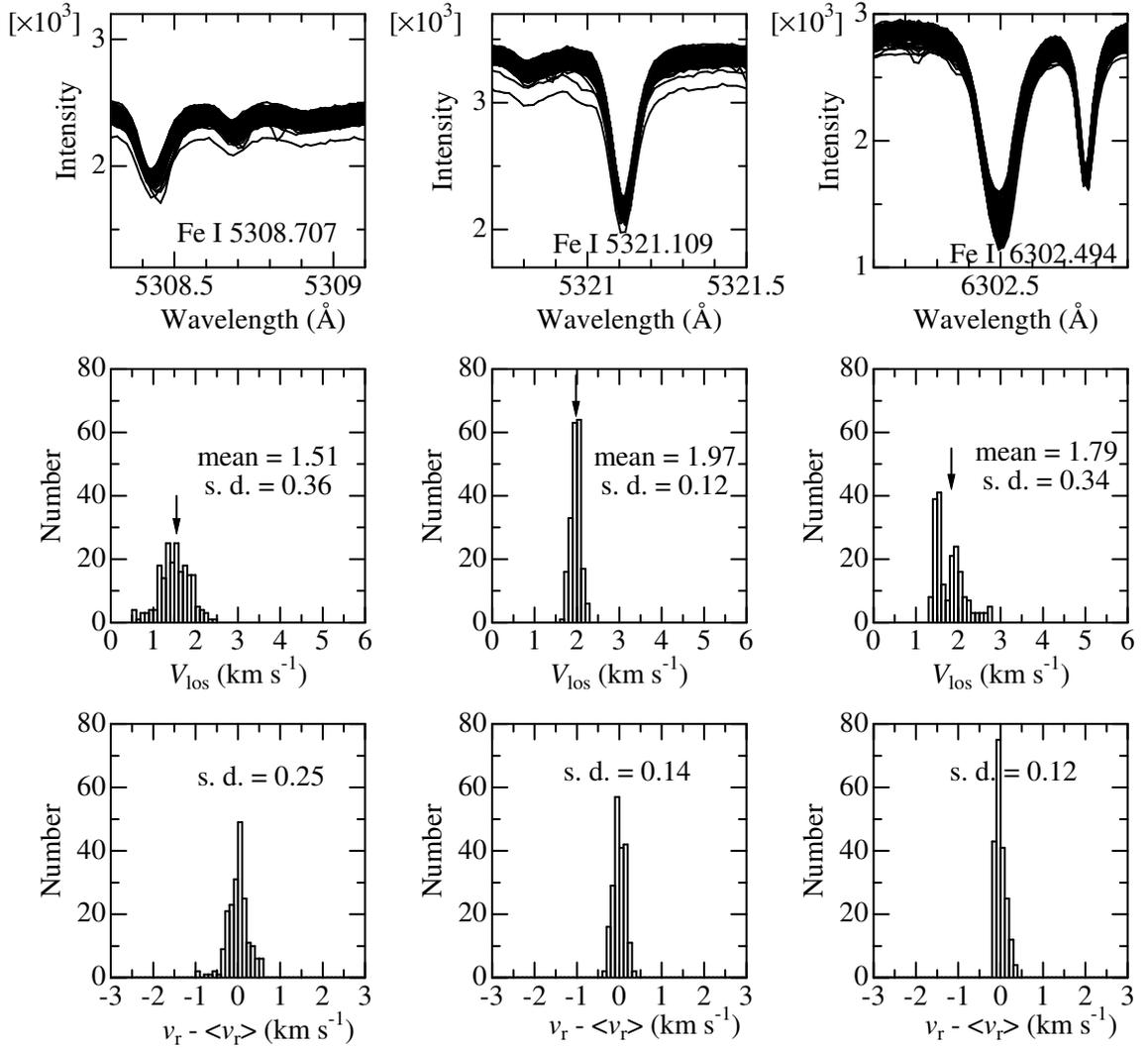}
  \end{center}
\caption{The upper panels show the 200 raw unnormalized Hida/DST 
spectra at each pixel (before being spatially averaged)
for the Fe~{\sc i} 5308.707 (left), Fe~{\sc i} 5321.709 (center), 
and Fe~{\sc i} 6302.494 (right) lines.
The histograms of $V_{\rm los}$ and $v_{\rm r} - \langle v_{\rm r} \rangle$
resulting from the analysis on these raw spectra are presented 
in the middle and lower panels, respectively (the mean value as well as 
the standard deviation of the distribution are also given).  
The downward arrows indicate the standard solution of $V_{\rm los}$ 
derived from the spatially averaged spectra.
}
\end{figure}

\begin{figure}
  \begin{center}
    \FigureFile(120mm,120mm){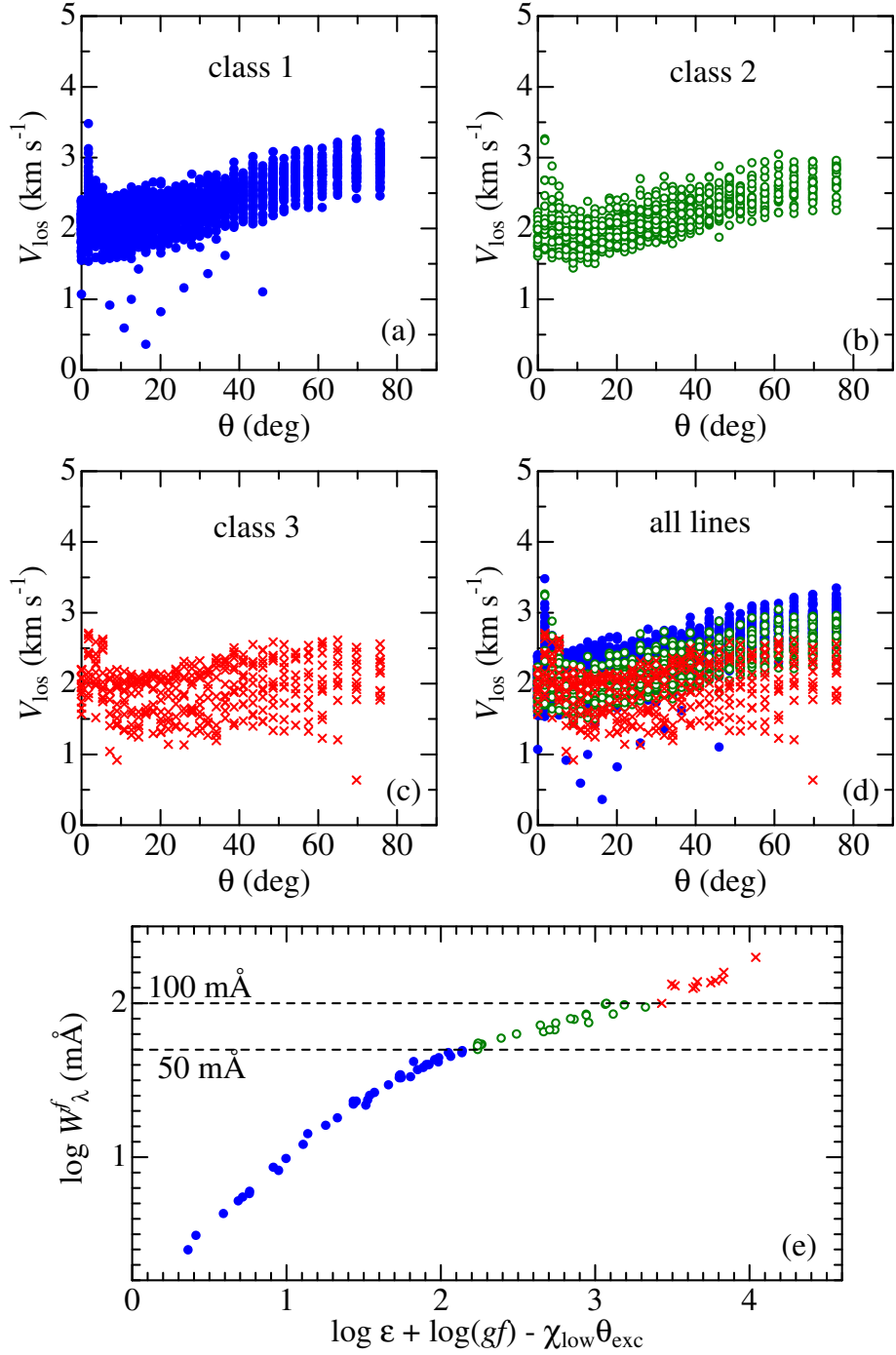}
  \end{center}
\caption{
Panels (a)--(d) show the $V_{\rm los}$ vs. $\theta$ diagram 
corresponding to each of the line-strength classes (cf. table~1),
which were derived from the analysis of Hida/DST data:
(a) unsaturated class-1 lines (filled circles), (b) moderately saturated 
class-2 lines (open circles), (c) strongly saturated class-3 lines
(crosses), and (d) all lines of class 1--3.
The curve of growth constructed by using the data of
Fe~{\sc i} lines presented in table~1 (on the assumption of $\xi = 0.5$~km~s$^{-1}$ 
and $\theta_{\rm exc}$ = $5040/T_{\rm exc} = 1$) is displayed in panel (e), 
which demonstrates the saturation degree of each line-strength class. 
}
\end{figure}

\begin{figure}
  \begin{center}
    \FigureFile(150mm,150mm){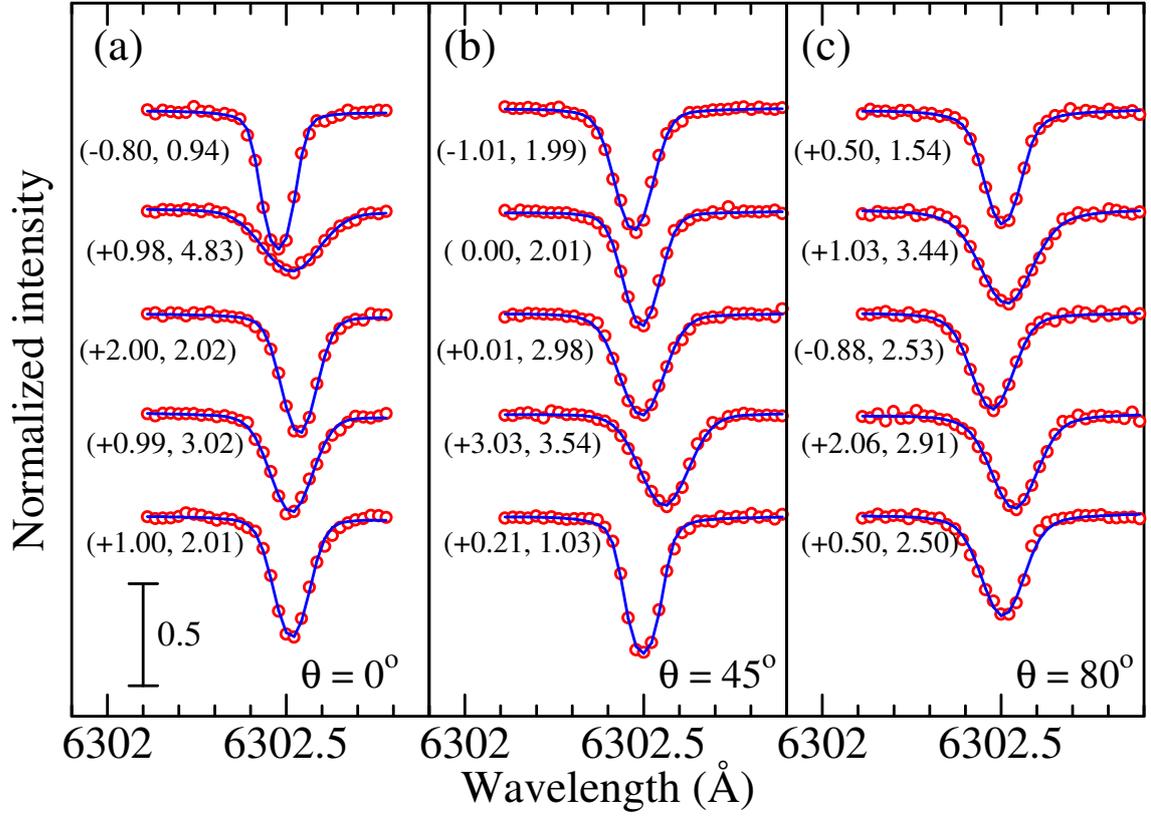}
  \end{center}
\caption{
Examples of Fe~{\sc i} 6302.494 line-profile fitting  
based on the {\it Hinode}/SOT data. Shown here are the selected 5 
representative cases of different ($v_{\rm r}$, $V_{\rm los}$) 
solutions indicated in the figure.
The observed spectra are shown in red circles while the best-fit
theoretical spectra are depicted in blue lines.
Panels (a), (b), and (c) correspond to $\theta = 0^{\circ}$, 
$\theta = 45^{\circ}$, and $\theta = 80^{\circ}$, respectively.
}
\end{figure}

\begin{figure}
  \begin{center}
    \FigureFile(160mm,160mm){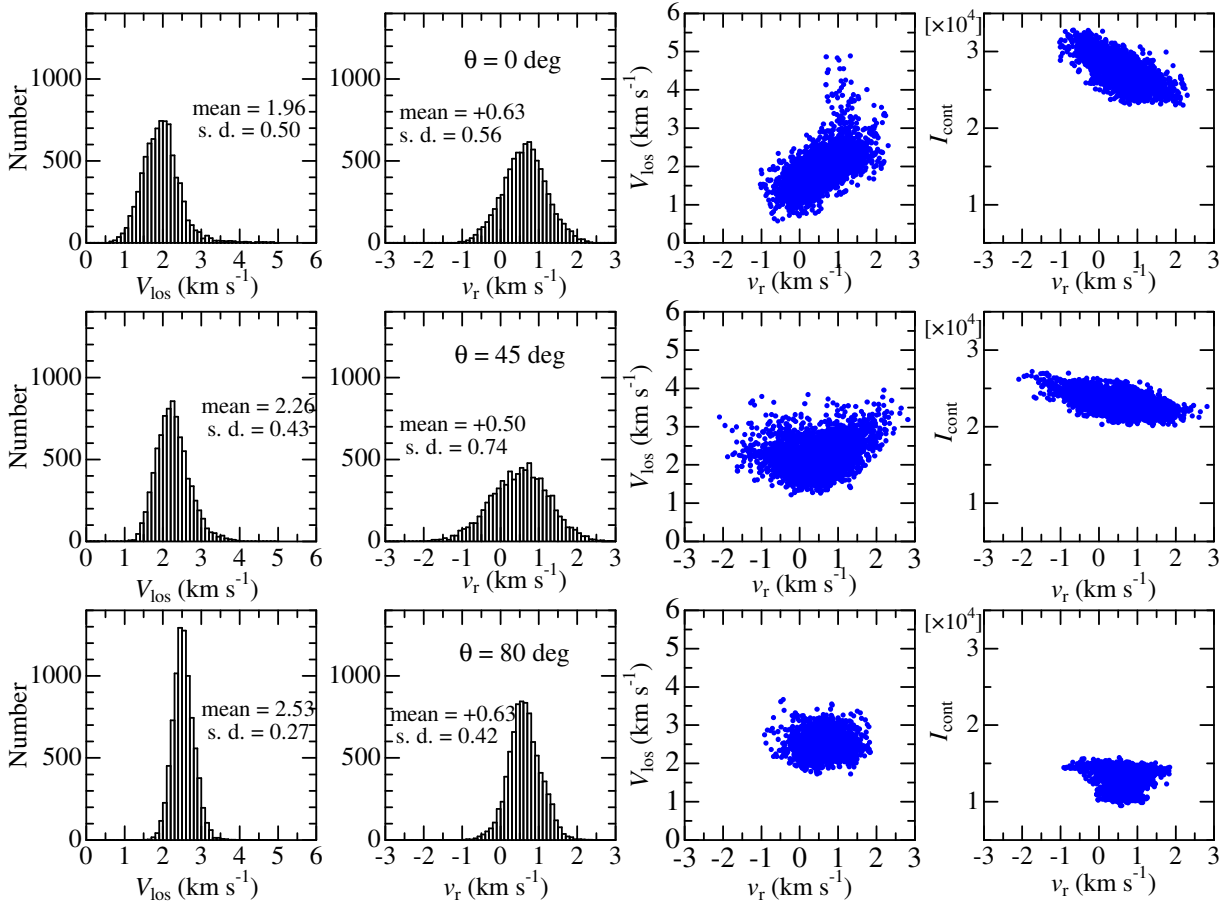}
  \end{center}
\caption{
Histograms and correlations of velocity parameters derived from the analysis of 
Hinode/SOT spectra. From left to right: Histogram of $V_{\rm los}$, 
histogram of $v_{\rm r}$, $V_{\rm los}$ vs. $v_{\rm r}$ correlation,
and $I_{\rm cont}$ (continuum intensity) vs. $v_{\rm r}$ correlation.
The mean value and the standard deviation are also indicated in each
histogram panel.
The upper, middle, and lower panels correspond to $\theta = 0^{\circ}$, 
$\theta = 45^{\circ}$, and $\theta = 80^{\circ}$, respectively
}
\end{figure}

\begin{figure}
  \begin{center}
    \FigureFile(150mm,170mm){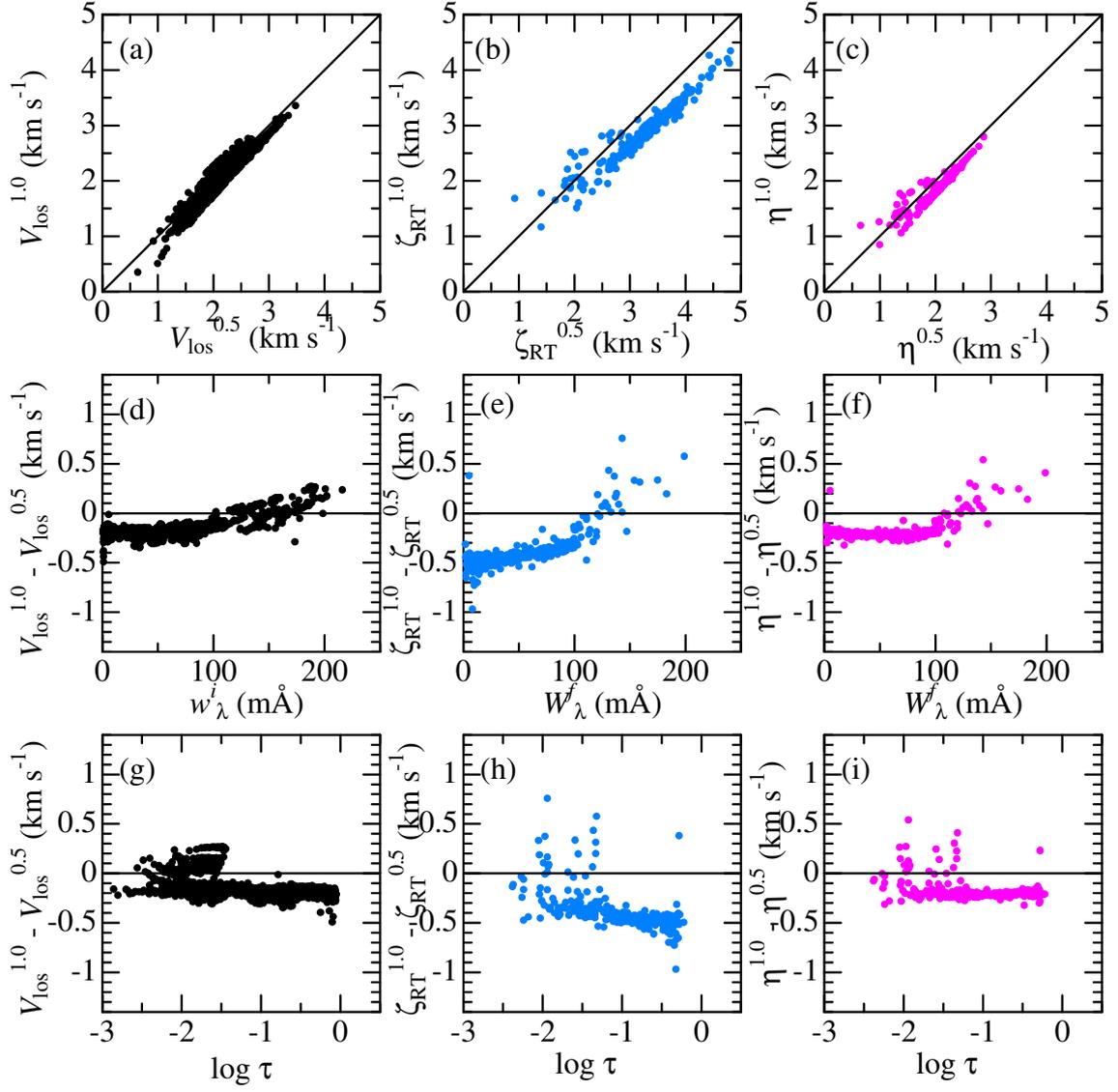}
  \end{center}
\caption{
Effect of changing the microturbulence ($\xi$) from the fiducial value of 0.5~km~s$^{-1}$
adopted in this study to 1.0~km~s$^{-1}$ on the results of $V_{\rm los}$ (left panels),
$\zeta_{\rm RT}$ (center panels), and $\eta$ (right panels) derived in section~1 
($\zeta_{\rm RT}$, $\eta$) and subsection~4.3 ($V_{\rm los}$).
Upper panels: Comparison of $\xi = 1.0$~km~s$^{-1}$ and $\xi = 0.5$~km~s$^{-1}$
results. Middle panels: Difference between $\xi = 1.0$~km~s$^{-1}$
and $\xi = 0.5$~km~s$^{-1}$ results plotted against the equivalent width ($w^{i}_{\lambda}$
in panel (d) is that derived from our intensity spectrum, while Meylan et al.'s (1993) 
values given in table~1 are adopted for $W^{f}_{\lambda}$ in panels (e) and (f)).
Lower panels: Difference between $\xi = 1.0$~km~s$^{-1}$ and $\xi = 0.5$~km~s$^{-1}$ 
results plotted against the mean formation depth.
}
\end{figure}

\begin{figure}
  \begin{center}
    \FigureFile(150mm,150mm){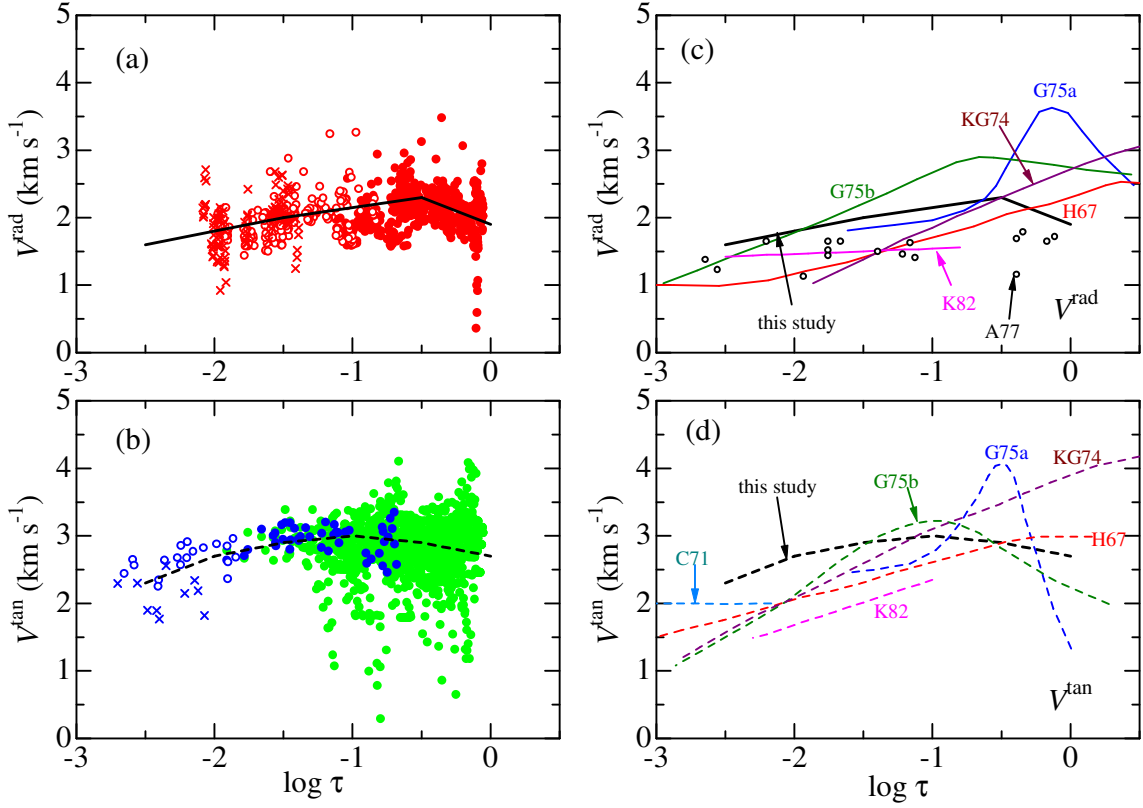}
  \end{center}
\caption{
The left two panels show the derivation of $V^{\rm rad}$ and $V^{\rm tan}$ 
(radial and tangential components of anisotropic Gaussian macroturbulence) 
based on the $V_{\rm los}$ data derived from our analysis of Hida/DST spectra.
(a): Results of $V_{\rm los}$ (line-of-sight 
velocity dispersion) at $1 \ge \cos \theta > 0.95$ ($0^{\circ} \le \theta < 17^{\circ}$), 
which we regard as $V^{\rm rad}$, plotted against $\langle \log \tau \rangle$,
where the thick solid line (drawn by eye-judgement) represents the 
approximate mean trend.  (b): The $V_{\rm los}$ values at $0.3 > \cos \theta$ 
($73^{\circ} < \theta$), which we regard as $V^{\rm tan}$,
are plotted against $\langle \log \tau \rangle$. 
The light green symbols show the $V^{\rm tan}$ results specially estimated 
from $V_{\rm los}$ at $0.95 > \cos \theta > 0.3$ ($17^{\circ} <\theta < 73^{\circ}$) 
by using equation~(A1) and the already derived mean $V^{\rm rad}(\tau)$ relation. 
The eye-judged trend of mean $V^{\rm tan}(\tau)$ is also shown (thick dashed line). 
In these two panels (a) and (b), the data derived from the line-strength class 
1, 2, and 3 (defined in table~1) are expressed in filled circles, open circles, 
and crosses, respectively.  
The right panels (c) and (d) summarize the various literature results regarding 
the depth-dependence of $V^{\rm rad}$ and $V^{\rm tan}$,
where the mean relations derived by ourselves are also overplotted by thick solid
and thick dashed lines, respectively.
Note that most of the literature data were read from Canfield and Beckers' (1976) 
figure~1 except for the data of Ayres (1977; open circles in panel (c)) and 
Kostik (1982).  A77 $\cdots$ Ayres (1977); C71 $\cdots$ Canfield (1971); 
G75a $\cdots$ Gurtovenko (1975a); G75b $\cdots$ Gurtovenko (1975b);
G75c $\cdots$ Gurtovenko (1975c); H67 $\cdots$ Holweger (1967); KG74 $\cdots$
Kondrashova and Gurtovenko (1974); K82 $\cdots$ Kostik (1982).
} 
\end{figure}

\begin{figure}
  \begin{center}
    \FigureFile(100mm,160mm){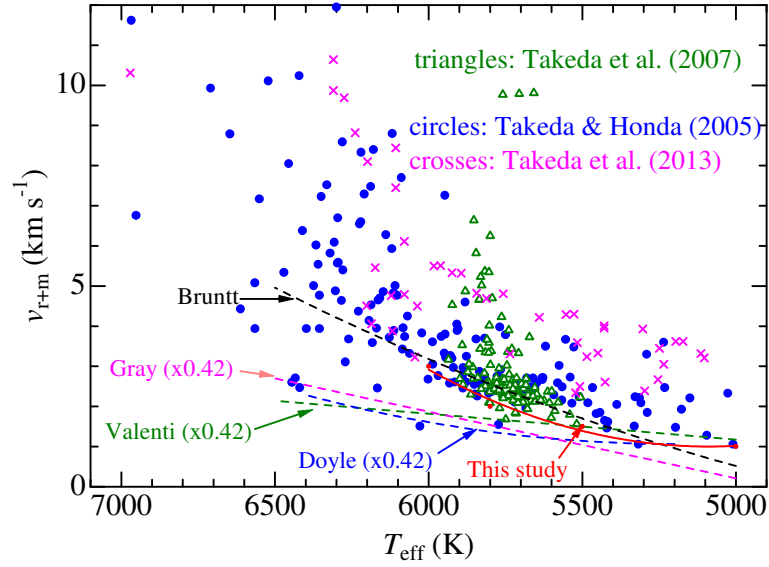}
  \end{center}
\caption{
Gaussian-approximated rotation plus macroturbulence broadening parameter
($v_{\rm r+m}$; derived from the spectrum-fitting analysis of 6080--6089~$\rm\AA$
region) plotted against $T_{\rm eff}$. Filled circles $\cdots$ FGK-type stars
(Takeda \& Honda 2005); open triangles $\cdots$ solar-analog stars (Takeda et al. 2007); 
crosses $\cdots$ Hyades stars of $T_{\rm eff} \le 6310$~K (Takeda et al. 2013).
Analytical relations of macroturbulence (reduced to the scale of Gaussian $v_{\rm mt}$) 
as function of $T_{\rm eff}$ proposed by previous studies (Gray 1984; 
Valenti \& Fischer 2005; Bruntt et al. 2010; Doyle et al. 2014) are shown 
by dashed lines, where the relation $v_{\rm mt} =  0.42 \zeta_{\rm RT}$
was applied to convert the $\zeta_{\rm RT}$ results (of Gray, Valenti \& Fischer, and 
Doyle et al.) into $v_{\rm mt}$ while Bruntt et al.'s Gaussian values were used unchanged.
The analytical relation for $v_{\rm mt}$ given by equation~(A3), which fits 
the envelope of this $v_{\rm r+m}$ distribution, is also depicted (solid line).
}
\end{figure}

\end{document}